\documentstyle[12pt]{article}
\makeatletter
\@addtoreset{equation}{subsection}
\makeatother
\def\pano{\par\noindent} \let\nn=\nonumber \def\hQ{\hat{Q}}
\def\homepage{\ $\tilde{\phantom\iota}$} \def\ZZ{\relax{\sf Z\kern-.4em Z}}
\def\inbar{\vrule height1.5ex width.4pt depth0pt}
\def\IC{\relax\,\hbox{$\inbar\kern-.3em{\rm C}$}}
\def\IN{\relax{\rm I\kern-.18em N}} \def\IP{\relax{\rm I\kern-.18em P}}
\def\fnote#1#2{\begingroup\def\thefootnote{#1}\footnote{#2}\addtocounter
{footnote}{-1}\endgroup}
\def\beq{\begin{equation}} \def\eeq{\end{equation}}
\def\bea{\begin{eqnarray}} \def\eea{\end{eqnarray}}
\def\lleq#1{\label{#1}\eeq} \def\llea#1{\label{#1}\eea}
\def\tabroom{\hbox to0pt{\phantom{\Huge A}\hss}}
\def\tab2{\hbox to0pt{\phantom{\huge A}\hss}}
\def\a{\alpha} \def\b{\beta} \def\k{\kappa} \def\th{\theta}
\def\g{\gamma} \def\l{\lambda} \def\si{\sigma} 
\def\G{\Gamma} \def\L{\Lambda} \def\Si{\Sigma} 
\def\bc{\bar c} \def\bl{\bar l} \def\bm{\bar m} \def\bq{\bar q}
\def\bs{\bar s} \def\bz{\bar z} \def\bJ{\bar J} \def\bQ{\bar Q}
\def\btau{\bar \tau} \def\bDelta{\bar \Delta} \def\bSi{\bar \Si}
\def\cA{{\cal A}} \def\cI{{\cal I}} \def\cO{{\cal O}} \def\cR{{\cal R}}
\def\vphi{\varphi} \def\tvphi{\tilde \vphi}
\def\vchi{\vec \chi} \def\vth{\vartheta}
\def\lra{\longrightarrow} \def\del{\partial}
\begin{document}
\baselineskip=18pt
\parskip=.1truein
\parindent=20pt
\hfill {hep--th/9510055}
\vskip -.1truein
\hfill {BONN--TH--95--17}
\vskip -.1truein
\hfill {IFP--601--UNC}
\vskip -.1truein
\hfill {NSF--ITP--95--120}
\vskip -.1truein
\hfill {October 1995}
\vskip .4truein
\centerline{\bf THE (0,2) EXACTLY SOLVABLE STRUCTURE OF CHIRAL RINGS,}
\vskip .1truein
\centerline{\bf LANDAU--GINZBURG THEORIES AND CALABI--YAU MANIFOLDS}
\vskip .4truein
\centerline{ R.Blumenhagen$^1$
              \fnote{\diamond}{Email:\ blumenha@physics.unc.edu},
             R.Schimmrigk$^{2,3}$
              \fnote{\dagger}{Email:\ netah@avzw02.physik.uni-bonn.de}
             and A.Wi\ss kirchen$^3$
              \fnote{\star}{Email:\ wisskirc@avzw02.physik.uni-bonn.de}}
\vskip .24truein
\centerline{\it $^1$ IFP, Department of Physics, University of North
 Carolina,}
\vskip .01truein
\centerline{\it Chapel Hill, NC 27599, USA}
\vskip .2truein
\centerline{\it $^2$ Institute for Theoretical Physics, University of
 California,}
\vskip .01truein
\centerline{\it Santa Barbara, CA 93106, USA}
\vskip .2truein
\centerline{\it $^3$ Physikalisches Institut der Universit\"at Bonn,}
\vskip .01truein
\centerline{\it Nu\ss allee 12, 53115 Bonn, Germany}
\vskip 0.4truein
\centerline{\bf ABSTRACT}
\noindent
We identify the exactly solvable theory of the conformal fixed point of
(0,2) Calabi--Yau $\si$--models and their Landau--Ginzburg phases. To this
end we consider a number of (0,2) models constructed from a particular
(2,2) exactly solvable theory via the method of simple currents. In order
to establish the relation between exactly solvable (0,2) vacua of the
heterotic string, (0,2) Landau--Ginzburg orbifolds and (0,2) Calabi--Yau
manifolds, we compute the Yukawa couplings of the exactly solvable model
and compare the results with the product structure of the chiral ring
which we extract from the structure of the massless spectrum of the exact
theory. We find complete agreement between the two up to a finite number
of renormalizations. For a particularly simple example we furthermore
derive the generating ideal of the chiral ring from a (0,2) linear
$\si$--model which has both a Landau--Ginzburg and a (0,2) Calabi--Yau
phase.
\renewcommand\thepage{}
\newpage
\pagenumbering{arabic}
\section{Introduction}
Ever since the revival of superstring theory ten years ago (2,2)
supersymmetric vacua have taken center stage. This has been the case
despite the fact that the main motivation for considering supersymmetric
groundstates, the hierarchy problem, actually necessitates only the
consideration of (0,2) conformal field theories \cite{bdfm88}. The reason
for this focus on an apparently rather special class of theories is to be
found not so much in the fact that (0,2) models were somewhat hard to come
by. Indeed, (0,2) models have been constructed a number of years ago by
several authors \cite{freeferm,lattice,asymorbi,fimqr89,sy90}
\fnote{1}{More recently (0,2) exactly solvable theories based on coset
      theories have been considered in \cite{bjkz95}.}.
More important is the fact that (2,2) theories are of phenomenological
interest, leading to three--generation models with small gauge groups at
the Planck scale \cite{rs87,rs90} and that they raise a number of
fundamental and intriguing questions. A short and incomplete list of such
interesting aspects might comprise the possibility of conifold phase
transitions \cite{cdls88,as95} in the early universe, the exact solution
of Calabi--Yau $\si$--models \cite{g88,g87,lg,ew93} as well as the fact
that the moduli space features a number of important properties, such as
special geometry \cite{as90}, world sheet mirror symmetry \cite{wms,ls90},
spacetime mirror symmetry \cite{sms,others} and scaling \cite{rs94}.
The full understanding of these problems has been slow to emerge and most
of them have not yet been resolved in a completely satisfactory fashion
even in the simpler context of (2,2) theories.
\par
The fact remains, however, that (0,2) theories do provide a
phenomenologically appealing framework \cite{ew86} and a tantalizing
question has been for some time what the generic features of the space of
(0,2) vacua are and which, if any, of the features mentioned above survive
in this more general context. Part of the problem has been that even
though exactly solvable (0,2) models have been known for many years, their
relation to (0,2) $\si$--models has remained obscure. Unlike the situation
in the framework of (2,2) compactifications, where the conformal fixed
points of particularly simple types of Calabi--Yau $\si$--models have been
shown \cite{g87,g88b,ss88,glr89,rs90}\cite{lr88,ls88,fkss89} to be
described by Gepner models \cite{g88} or Kazama--Suzuki models
\cite{ks89,lvw89,ls91,ans91}, no such understanding has emerged in the
context of (0,2) theories. It is this question of the existence and nature
of the exact theory describing the conformal fixed points of (0,2)
$\si$--models which we address in the present paper.
\par
In \cite{ew93} Witten formulated a framework which has been employed in
\cite{dk94} to formulate a class of (0,2) supersymmetric Landau--Ginzburg
models, generalizing the construction of the class of (2,2)
Landau--Ginzburg theories \cite{lg,kw93}. This Landau--Ginzburg
formulation allows us to address the problem of destabilization of (0,2)
$\si$--models \cite{dsww86,egnq86,jd87,dg88,dk94b,sw95} and facilitates
the computation of some important characteristics of these theories,
but falls short of addressing the important problem of identifying
the exactly solvable structure of the superconformal fixed points.
In \cite{bw95} a class of exactly solvable (0,2) models with
$(c,\bc)=(6+r, 9)$, $r=3,4,5$ and gauge group of rank $(9-r)$ was
constructed by generating new modular invariants from
the class of Gepner models via a slight modification of the simple
current method of \cite{sy89,sy90}. By considering simple currents which
break part of the supersymmetry one finds that in general there exist
several (0,2) daughter theories which can be built from any of the Gepner
models. The resulting class of theories thus is much richer than the
original class of (2,2) tensor models. Furthermore, even for theories with
an E$_6$ gauge group one obtains spectra which do not appear among the
class of Gepner models \cite{g88}, the complete list of which has been
constructed in \cite{lr88,ls88,fkss89}
\fnote{2}{They do appear in the class of all Landau--Ginzburg theories
          \cite{ks94,krsk92} and therefore might describe (0,2)
          deformations of known (2,2) theories. The complete list of
          Landau--Ginzburg theories can be accessed on the web at the
          Calabi--Yau pages at \\
 {\it http://www.math.okstate.edu/{\homepage}katz/CY}\\
           and its European mirror \\
 {\it http://thew02.physik.uni-bonn.de/{\homepage}netah/cy.html}. \\
   (These pages are in an experimental stage and prone to changes.)}.
This supports the expectation that (2,2) theories describe but a small
part of the total space of all vacua with N$=$1 spacetime supersymmetry.
\par
In the present paper we generalize the (2,2) triality of exactly
solvable models, Landau--Ginzburg theories and Calabi--Yau manifolds, 
to the context of (0,2) string vacua by establishing a relation
between the linear $\si$--models considered in \cite{dk94} and the
exactly solvable models of \cite{bw95}. The technique we use is
reminiscent of the construction introduced in \cite{ls90}. It was shown
there that for a particular class of discrete symmetries acting on a
(2,2) exactly solvable model or its Landau--Ginzburg counterpart,
it is possible not only to derive the anomalous dimensions of the
chiral primary fields of the orbifold, but also to find a
superpotential which describes the orbifold theory.
The implementation of simple currents is similar to the action of
discrete symmetries in that they also change the dimensions of the
original fields. We show that by considering the detailed structure of
the massless spectrum it is possible to extract the anomalous
dimensions of the scaling fields of the (0,2) theory and to derive the
chiral ring describing the (0,2) theory. We furthermore show that our
identification is correct by computing the Yukawa couplings
of the exactly solvable models as well as the product structure of
the chiral ring. Comparing the two emerging patterns leads to
perfect agreement up to a change of basis.
\par
The paper is organized as follows. In Section 2 we review the basic
ingredients of the construction of \cite{bw95} before we proceed
in Sections 3 and 4 to establish in some detail the relation between
a particularly simple exactly solvable (0,2) model
and a linear (0,2) $\si$--model. To this end we first determine in
Section 3 the massless spectrum and compute the Yukawa couplings in the
exact theory. In Section 4 we derive the chiral ring from the structure
of the generations of the exact model and analyze its product structure,
the comparison of which with the exact Yukawa couplings leads to
complete agreement. We close this Section by showing that the resulting
chiral ring can be derived from a particular (0,2) linear $\si$--model
defining a stable bundle over a smooth Calabi--Yau manifold. This linear
$\si$--model features both a Landau--Ginzburg phase and a (0,2) Calabi--Yau
phase and therefore the exactly solvable model described in Section 3
indeed provides the underlying exact conformal field theory
of a (0,2) Calabi--Yau manifold at a particular point in the moduli space. 
It then follows from the work of \cite{dk94b,sw95} that the theories 
contained in a neighborhood of this point in moduli space are also 
conformally invariant.
In the remaining Sections we extend our considerations to further models.
\section{Exactly solvable theories}
\vskip -.1truein
\subsection{New models from old via simple currents}
The class of exactly solvable models which we will focus on has been
described in some detail in \cite{bw95}. In order to make the
present article self--contained we begin by reviewing the salient
features of the construction.
\par
The basic tool for the construction is the simple current technique
\cite{sy89,sy90} for building new modular invariants from old ones.
Briefly, one considers a rational conformal field theory with a given
modular invariant partition function which is supposed to contain a simple
current J of index $N$ and monodromy parameter $R$, i.e.\ a unipotent
field (J$^N=1$) such that for primary fields $\Phi_i$ of the theory
  \beq
   {\rm J} \times \Phi_i = \Phi_j,~~~~
  \Delta ({\rm J}) = \frac{R(N-1)}{2N}~~{\rm mod}~~1.
   \eeq
$R$ then is defined modulo $N$ for $N$ odd and modulo $2N$ for $N$ even.
Furthermore one introduces a monodromy charge $Q_{\rm J}$ for
primary fields associated with a given simple current J
\beq
    Q_{\rm J}(\Phi_i)=\Delta(\Phi_i)+\Delta({\rm J}) -
      \Delta({\rm J} \cdot \Phi_i) ~~ {\rm mod}~~1,
\eeq
which takes values $\frac{t}{N}$, $t\in \ZZ$.
Of importance to the construction will be a slightly modified monodromy
charge defined on the element of each orbit by
\beq
      \hQ({\rm J}^p \Phi_i) = \frac{t+pR}{2N}~~{\rm mod}~~1.
\eeq
The simple current and its iterative application defines orbits of all the
primary fields $\Phi_i$, J $\Phi_i$, $\cdots ,$ J$^d \Phi_i$,
where $d$ is a divisor of $N$. If $R$ is even the matrix
\beq
  M_{ij}({\rm J}):=\delta^1\left(\hQ (\Phi_i) + \hQ (\Phi_j)\right)
                  \sum_{p=1}^{\rm N} \delta(\Phi_i, {\rm J}^p\Phi_j) ~
\eeq
with $\delta^1(x)=1$ for $x\in \ZZ$ and zero otherwise, defines
a new modular invariant partition function
\beq
        Z(\tau, \btau) = \sum_{i,j} \chi_i(\tau) M_{ij} \chi_j(\btau).
\eeq
This procedure allows for an iteration procedure with a whole bunch of
simple currents by considering
\beq
Z(\tau,\btau) \sim \vchi(\tau) M({\rm J}_n) \cdots M({\rm J}_2)
     M({\rm J}_1) \vchi(\btau),
\eeq
where $\sim$ indicates equality up to an overall factor originating
from a universal multiplicity factor.
\vskip -.1truein
\subsection{(0,2) Simple current modular invariants leading to E$_{9-r}$}
We now wish to apply the above general considerations to the particular
case, where the final theory we end up with has the following properties:\
a) The gauge group is any of the groups
    E$_{9-r} \ni \{{\rm E}_6, {\rm E}_5={\rm SO(10)},
     {\rm E}_4={\rm SU(5)}, {\rm E}_3={\rm SU(3)}\times {\rm SU(2)}\}$.
b) The central charges in the two internal sectors are $(c,\bc)=(6+r,9)$.
c) Besides the right moving U(1)$_{\rm R}$ current which is part of the
   right moving N$=$2 superconformal theory, there exists also a left
   moving U(1)$_{\rm L}$ current J$_{\rm L}$ satisfying the
   operator product expansion
\beq
   {\rm J}_{\rm L}(z) {\rm J}_{\rm L}(w) = \frac{r}{(z-w)^2} + {\rm reg.}
\lleq{so10-ope}
d) Only the subset SO(16$-2r$)$\times $ U(1)$_{\rm L} \subset $ E$_{9-r}$
   of the gauge group is linearly realized, the full E$_{9-r}$
   being generated by taking orbits with
   respect to the spectral flow operator of conformal dimension
   $(\Delta, Q)=(\frac{r}{8},\frac{r}{2})$.
It is convenient to describe the construction in the following left--right
symmetric way, where the asymmetry between the left and right sector is
achieved at the end of the day by throwing away part of the right moving
current algebra in such a way as not to endanger modular invariance.
Consider then an internal conformal field theory with the ingredients
in Table 1.
\begin{center}
\begin{tabular}{|l r r |}
\hline
             &Left Sector $c$  &Right Sector $\bc$  \tabroom  \\
\hline
N$=$2 SCFT   &9            &9            \tabroom \\
$\left({\rm  U(1)}_2\right)^{r-3}$
             &$r-3$        &$r-3$  \tabroom \\
SO(16$-2r)\times $E$_8$
             &$16-r$       &$16-r$\tabroom \\
\hline
\end{tabular}
\end{center}
\centerline{{\bf Table 1:}~{\it Ingredients for the construction of the
                        internal theory.}}
\par
The crucial, new ingredient, as compared to Gepner's tensor model
construction, is the free boson $\phi$ compactified on a circle
of radius R$=$2, denoted by U(1)$_2$. The diagonal partition function
of this theory is simply
\beq
Z_{{\rm U(1)}_2}(\tau, \btau)
= \sum_{m=-1}^2 \Theta_{m,2}(\tau) \Theta_{m,2}(\btau).
\eeq
The conformal dimension and charge of the primary
fields $\Phi^{{\rm U(1)}_2}_{m,2}$ are
\beq
\left(\Delta, Q\right)(\Phi^{{\rm U(1)}_2}_{m,2}) =
  \left(\frac{m^2}{8}, \frac{m}{2}\right).
\eeq
The current $j_{{\rm U(1)}_2} = i \del_z \phi$ satisfies the following OPE
\beq
j_{{\rm U(1)}_2}(z) j_{{\rm U(1)}_2}(w) =  \frac{1}{(z-w)^2} + {\rm reg.}
\eeq
\par
Consider first the left moving sector. Even though the U(1)$_2$
theory is not N$=$2 supersymmetric, it does feature a spectral flow
between the $m$ even sectors and the $m$ odd sectors. Since this
boson describes a Dirac fermion the spectral flow operator
$ e^{i\phi(z)/2}$, with conformal dimension and charge
$(\frac{1}{8}, \frac{1}{2})$, relates the NS sector and
the R sector of the Dirac fermion. The introduction of the U(1)$_2$
current algebra is motivated by the fact that by combining its current
$j_{{\rm U(1)}_2}$ with the U(1) current
$j_{c=9} = i \sqrt{3}\del \Phi$ of the N$=$2, c$=$9 theory one obtains
the OPE (\ref{so10-ope}) for U(1)$_{\rm L}$. Furthermore, putting together
the spectral flow operators of the N$=$2 conformal field theory with
the U(1)$_2$ one obtains the spectral flow operator of the
$c=(6+r)$ left moving sector
\beq
\Si_{c=6+r}(z) = e^{i\frac{\sqrt{3}}{2} \Phi(z)} \cdot
    \prod_{j=1}^{r-3} e^{\frac{i}{2} \phi_j(z)}.
\eeq
It can be shown that this operator does indeed extend the group
SO(16$-2r$)$\times $ U(1)$_{\rm L}$ to E$_{9-r}$.
\par
Before turning to the right moving sector, it is useful to review
a few facts about the representations of SO($2n$) Kac--Moody algebras
at level $k=1$. Recall that the representations of an SO(2$n$) algebra
decompose into scalars~(0), spinors~$(s)$, antispinors~$(c)$ and
vectors~$(v)$. The characters and the quantum numbers of the
corresponding primary fields are collected in Table 2.
\begin{center}
\begin{tabular}{| l c c c |}
\hline
Character &Conformal Dimension $\Delta$ &$Q$ mod 2 &Degeneracy \tabroom \\
\hline
$\chi_0 = \frac{1}{2} \left(\left(\frac{\vth_3}{\eta}\right)^n +
                            \left(\frac{\vth_4}{\eta}\right)^n \right)$
           & 0          & 0      &1    \tabroom \\
$\chi_v = \frac{1}{2} \left(\left(\frac{\vth_3}{\eta}\right)^n -
                            \left(\frac{\vth_4}{\eta}\right)^n \right)$
           &$\frac{1}{2}$  &1      &$2n$    \tabroom \\
$\chi_s = \frac{1}{2} \left(\left(\frac{\vth_2}{\eta}\right)^n +
                            \left(\frac{\vth_1}{\eta}\right)^n \right)$
           &$\frac{n}{8}$ &$\frac{n}{2}$     &$2^{n-1}$    \tabroom \\
$\chi_c = \frac{1}{2} \left(\left(\frac{\vth_2}{\eta}\right)^n -
                            \left(\frac{\vth_1}{\eta}\right)^n \right)$
           &$\frac{n}{8}$  &$\frac{n}{2}-1$  &$2^{n-1}$ \tabroom \\ [1ex]
\hline
\end{tabular}
\end{center}
\centerline{{\bf Table 2:}~{\it Characters of SO(2n). The charge
       $Q$ is taken with respect to the sum of all}}
\centerline{\it ~~~~~~~~~~~~~Cartan elements of the
       Lie algebra and $\vth_i$ denote the Jacobi theta functions.}
\par
Because our goal is to eventually turn the bosonic theory into a heterotic
theory by applying the bosonic string map as considered by Gepner
\cite{g88}
\bea
&\chi_0^{{\rm SO(10)}\times {\rm E}_8} \lra \chi_v^{\rm SO(2)}
&~~~\chi_v^{{\rm SO(10)}\times {\rm E}_8} \lra \chi_0^{\rm SO(2)} \nn \\
&\chi_s^{{\rm SO(10)}\times {\rm E}_8} \lra -\chi_c^{\rm SO(2)}
&~~~\chi_c^{{\rm SO(10)}\times {\rm E}_8} \lra -\chi_s^{\rm SO(2)}, 
\eea
we wish to extend the SO(16$-2r)\times $U(1)$_2^{r-3}$ to SO(10). In the
right sector this is achieved by considering $(r-3)$ simple currents of
the form
\beq
{\rm J}_{\rm ext}
=\Phi^{{\rm U(1)}_2}_{2,2} \otimes \Phi_v^{{\rm SO(16}-2r)},
\eeq
with $\Delta({\rm J}_{\rm ext})=1$, which generate orbits of SO(10).
For $r=4$, say, these orbits take the form
\bea
& \chi_0^{\rm SO(10)}=
     \chi_0^{\rm SO(8)}\Theta_{0,2}+\chi_v^{\rm SO(8)}\Theta_{2,2}
&~~~\chi_v^{\rm SO(10)}=
     \chi_0^{\rm SO(8)}\Theta_{2,2}+\chi_v^{\rm SO(8)}\Theta_{0,2} \nn \\
&\chi_s^{\rm SO(10)}=
      \chi_s^{\rm SO(8)}\Theta_{1,2}+\chi_c^{\rm SO(8)}\Theta_{-1,2}
&~~~\chi_c^{\rm SO(10)}=
    \chi_c^{\rm SO(8)}\Theta_{0,2}+\chi_s^{\rm SO(8)}\Theta_{2,2}.
\eea
So far our considerations have been completely general. From this point
on we focus our attention on products of N$=$2 minimal tensor models a
la Gepner.
\vskip -.1truein
\subsection{(0,2) Theories from Gepner models}
Recall that for the minimal models the conformal anomaly is
\beq
c= \frac{3k}{k+2}
\eeq
and the dimensions and charges of the chiral primary fields
$\Phi^{l,q,s}$ are given by
\beq
\Delta^{l,q,s} = \frac{l(l+2)-q^2}{4(k+2)} + \frac{s^2}{8}, ~~~~~~
Q^{l,q,s} = -\frac{q}{k+2} + \frac{s}{2},
\eeq
where the level $k\in \IN$, $0\leq l \leq k$, $l-q+s \in 2\ZZ$ and
$|q-s|\leq l$. Here the range of the various quantum numbers is
$l=0,...,k$, $q\sim q+2(k+2)$, $s\sim s+4$ and we will employ the
following notation for the complete fields
\beq
\Phi^{l,q,s}_{\bl,\bq,\bs}=\left[\matrix{l&q&s\cr \bl&\bq&\bs\cr}\right],
\eeq
which, if $l = \bl$, simplifies into
\beq
\Phi^{l,q,s}_{l,\bq,\bs} = \left[l~\matrix{q&s\cr \bq&\bs\cr}\right].
\eeq
Important, finally, is the identification
\beq
\left[\matrix{l&q&s\cr \bl &\bq &\bs\cr}\right]
\sim \left[\matrix{k-l&q+(k+2)&s+2\cr k-\bl &\bq+(k+2) &\bs+2\cr}\right].
\eeq
\par
In \cite{bw95} the tensor models $\bigotimes_{i=1}^n k_i$ have been used
to construct (0,2) theories with the aforementioned properties a) -- d)
by breaking the left moving supersymmetry as well as the E$_6$ gauge
group present in Gepner's construction. The following steps do the job:
\begin{itemize}
\item The simple currents extending SO(16$-2r)\times $U(1)$_2^{r-3}$
    to SO(10) in the tensor model can be written as
  \beq
    {\rm J}^j_{\rm ext} = [0~0~0]^{\otimes n} \otimes
         [0]^{j-1}[2][0]^{r-3-j}\otimes [v],~~j=1,...,r-3,
  \eeq
  where $[v]$ denotes the vector representation of  SO(16$-2r$).
\item Next one has to make sure, as usual, that only fields of same
    character couple, i.e.\ Neveu--Schwarz ($s\in \{0,2\}$) to
    Neveu--Schwarz and Ramond ($s \in \{1,3\}$) to Ramond. This is
    achieved via the projection operators
    ${\rm J}_i = G_i \otimes \Phi_v^{{\rm SO}(16-2r)}$, where the
     $G_i= \Phi^{k,k+2,4}$ are the supercurrents
     in the $i$'th factor of the tensor model. The appropriate fields in
     a tensor model with $n$ factors take the (chiral) form
    \beq
     {\rm J}_i = [0~0~0]^{\otimes (i-1)} \otimes [k~(k+2)~4]
                \otimes [0~0~0]^{\otimes (n-i)}\otimes [0]^{r-3}
             \otimes [v],~~i=1,...,n.
    \eeq
\item The next step is to implement the right moving GSO projection
    onto states with even overall charge. This is achieved via the
     simple current
\beq
{\rm \bJ}_{\rm GSO} = \bSi_{c=9} \otimes
              \left[\matrix{0\cr 1\cr}\right]^{\otimes (r-3)}
             \otimes \left[\matrix{0\cr s\cr}\right],
\eeq
where
\beq
\bSi_{c=9} = \left[0~\matrix{0&0\cr 1&1\cr}\right]^{\otimes n}
\lleq{right-susy}
is the right moving spectral flow operator of dimensions
$(\Delta, \bDelta)(\bSi_{c=9})  = (0, \frac{3}{8})$ and charges
$(Q, \bQ)(\bSi_{c=9}) =(0, \frac{3}{2})$.
Implementing these projections leads to the partition function
\beq
Z \sim \vchi(\tau) M({\rm \bJ}_{\rm GSO})
      \left(\prod_{i=1}^n M({\rm J}_i)\right)
      \left(\prod_{j=1}^{r-3}M({\rm J}^j_{\rm ext})\right) \vchi(\btau).
\lleq{dor-pf}
At this point all the conditions for a (2,2) supersymmetric theory
have been implemented. After turning this left--right symmetric
theory into a class of heterotic string vacua by applying the
bosonic string map in the way described above, we should expect
the procedure described thus far to provide an alternative construction
of Gepner's models. The fact that the partition function
(\ref{dor-pf}) indeed reproduces the expected spectra provides
a nice check of the implementation.
\item
Our goal, however, was to obtain an asymmetric (0,2) CFT and
in order to do so one simply has to introduce further simple currents
in the left moving sector, denoted by $\Upsilon_i$ in the following,
which do not commute with the simple currents that appear in
(\ref{dor-pf}). Of particular interest are simple currents
$\Upsilon_i$ which break both the left moving N$=$2 supersymmetry and
the E$_6$ gauge group which results from the ${\rm \bJ}_{\rm GSO}$
projection. It turns out that for each tensor model one can find a
multitude of such simple currents, making this class an interesting
framework. Given the existence of fields $\Upsilon_i$ with the
appropriate properties the only remaining ingredient is the left
moving GSO projection which is implemented by the simple current
\beq
{\rm J}_{\rm GSO}=\Si_{c=6+r} \otimes \left[\matrix{s\cr 0\cr}\right],
\eeq
where
\beq
\Si_{c=6+r} = \left[0~\matrix{1&1\cr 0&0\cr}\right]^{\otimes n} \otimes
     \left[\matrix{1\cr 0\cr}\right]^{\otimes (r-3)}.
\eeq
The crucial effect of the simple currents $\Upsilon_i$ is that they
prevent the extension of the gauge group to E$_6$ via the currents
J$_i$ and J$_{\rm ext}$. Instead the chiral spectral flow is
$\Si_{c=6+r}^2$ and thus is an operator of dimension
$(\Delta, \bDelta) = (\frac{r}{2},0)$ and charges $(Q, \bQ)=(r,0)$.
It is this operator which extends the SO(16$-2r)\times $U(1)$_{\rm L}$
to E$_{9-r}$.
\end{itemize}
\par
Putting everything together then leads to the final form of the
partition function \cite{bw95}
\beq
Z\sim \vchi(\tau) M({\rm J}_{\rm GSO}) \left(\prod_i M(\Upsilon_i)\right)
       M({\rm \bJ}_{\rm GSO}) \left(\prod_{i=1}^r M({\rm J}_i)\right)
  \left(\prod_{j=1}^{r-3} M({\rm J}^j_{\rm ext})\right) \vchi(\btau).
\lleq{02pf}
This partition function exhibits all the desired features in order to
be of use in the exploration of exactly solvable models which possibly
describe particular points in the moduli space of Landau--Ginzburg
theories constructed in \cite{dk94}. In short one might summarize the
structure of these vacua as described in Table 3.
\pano
\begin{small}
\begin{center}
\begin{tabular}{|l r r |}
\hline
                       &Left Sector       &Right Sector  \tab2  \\
\hline
Ghosts $b,c$           &$-26$             &$-26$      \tabroom  \\
Super Ghosts $\b, \g$  &--                &$ 11$      \tabroom \\
Spacetime Coordinates $X^{\mu}$
                       &4                 &4          \tabroom \\
Superpartners $\psi^{\mu}$ of $X^{\mu}$
                       &--                &2  \tabroom \\
Internal CFT           &6+$r$             &9            \tabroom \\
Gauge Group  SO(16$-2r$)$\times $E$_8$
                       &16$-r$            &--          \tabroom \\
\hline
\end{tabular}
\end{center}
\end{small}
\centerline{{\bf Table 3:}~{\it Anomaly structure of the
                 complete (0,2) theory.}}
\section{An (80,0) SO(10) (0,2)--model}
The exact theories we focus on in the following are all derived from
the parent `quintic' tensor model defined by considering the product
of five N$=$2 minimal factors at level $k$$=$3. We will denote this
theory by $3^{\otimes 5}$. The different spectra and gauge groups
will be obtained by applying various simple currents $\Upsilon_i$.
\vskip -.1truein
\subsection{The massless spectrum}
Our first exact model is based on the simple current
\beq
\Upsilon_1=[3~0~-1]\otimes [0~0~0]^{\otimes 4}\otimes [1]\otimes [0],
\eeq
which only affects the first of the five minimal
N$=$2 factors and turns out to break the E$_6$ of the parent Gepner model
down to SO(10). The spectrum therefore is arranged into representations
of SO(10) and we have summarized the relevant multiplicities of the
massless sector in Table 4.
\pano
\begin{small}
\begin{center}
\begin{tabular}{|l c c c c|}
\hline
SO(10) Representation: &{\bf 0} &{\bf 10} &{\bf 16} &
 ${\bf {\overline {16}}}$  \tabroom \\
\hline
Spin 0:         & 350      &74        &80        &0          \tabroom \\
Spin 1:         & 7        & 0         &0        &0          \tabroom \\
\hline
\end{tabular}
\end{center}
\end{small}
\centerline{{\bf Table 4:}~{\it Massless spectrum of the
         (80,0) SO(10) daughter of the $3^{\otimes 5}$ model.}}
\par
Of particular relevance for the following are the generations, which
can be represented in a number of ways. Considering the
internal part of a generation only, the {\bf 16} of the SO(10) in the
(--1) ghost picture as a spacetime scalar leads to a vertex operator of the
form (at zero momentum)
\beq
V_{-1}^a = e^{-\rho (\bz) } \cO_{16} (z,\bz) \l^a,
\eeq
where the internal operator $ \cO_{16} (z,\bz)$ has U(1) charge
$(Q,\bQ) =(-1,-1)$,
and $\rho (\bz) $ is the bosonized supersymmetry ghost, whereas the
fermions $\l^a$ generate the SO(10) current algebra. The operator has
$\Delta = \bDelta =\frac{1}{2}$ and therefore defines a chiral primary
field. In this representation the generations take the form exhibited in
Table 5, in which we have used the abbreviations
\beq
g_0 = \left[2~\matrix{2&0\cr 2&0\cr}\right],~~~
g_1 = \left[2~\matrix{-3&-1\cr 2&0\cr}\right]
\lleq{gen-abbr}
for the two states that appear in the first minimal factor and
\beq
u_0 = \left[\matrix{0\cr 0\cr}\right],~~~
u_1 = \left[\matrix{-1\cr 0\cr}\right]
\lleq{u1-abbr}
for the U(1)--part of the generations.
\pano
\begin{scriptsize}
\begin{center}
\begin{tabular}{| l r r |}
\hline
Type   &Field      &Number \tabroom \\
\hline
I &$\left[0~\matrix{0&0\cr 0&0\cr}\right]
     \left[0~\matrix{0&0\cr 0&0\cr}\right]
    \left[0~\matrix{0&0\cr 0&0\cr}\right]
     \left[2~\matrix{2&0\cr 2&0\cr}\right]
    \left[3~\matrix{3&0\cr 3&0\cr}\right]
        ~u_0~ \left[\matrix{v\cr v\cr}\right]$ &12 \tabroom \\
II &$\left[0~\matrix{0&0\cr 0&0\cr}\right]
      \left[0~\matrix{0&0\cr 0&0\cr}\right]
     \left[1~\matrix{1&0\cr 1&0\cr}\right]
      \left[1~\matrix{1&0\cr 1&0\cr}\right]
     \left[3~\matrix{3&0\cr 3&0\cr}\right]
      ~u_0~ \left[\matrix{v\cr v\cr}\right]$ &12 \tabroom \\
III&$\left[0~\matrix{0&0\cr 0&0\cr}\right]
      \left[0~\matrix{0&0\cr 0&0\cr}\right]
     \left[1~\matrix{1&0\cr 1&0\cr}\right]
      \left[2~\matrix{2&0\cr 2&0\cr}\right]
      \left[2~\matrix{2&0\cr 2&0\cr}\right]
      ~u_0~ \left[\matrix{v\cr v\cr}\right]$ &12 \tabroom \\
IV &$\left[0~\matrix{0&0\cr 0&0\cr}\right]
      \left[1~\matrix{1&0\cr 1&0\cr}\right]
     \left[1~\matrix{1&0\cr 1&0\cr}\right]
      \left[1~\matrix{1&0\cr 1&0\cr}\right]
     \left[2~\matrix{2&0\cr 2&0\cr}\right]
     ~u_0~ \left[\matrix{v\cr v\cr}\right]$ &4 \tabroom \\
V$_i$ &$ g_i~
    \left[0~\matrix{0&0\cr 0&0\cr}\right]
     \left[0~\matrix{0&0\cr 0&0\cr}\right]
    \left[0~\matrix{0&0\cr 0&0\cr}\right]
     \left[3~\matrix{3&0\cr 3&0\cr}\right]
     ~u_i~ \left[\matrix{v\cr v\cr}\right]$ &8 \tabroom \\
VI$_i$ &$g_i~
      \left[0~\matrix{0&0\cr 0&0\cr}\right]
         \left[0~\matrix{0&0\cr 0&0\cr}\right]
      \left[1~\matrix{1&0\cr 1&0\cr}\right]
         \left[2~\matrix{2&0\cr 2&0\cr}\right]
         ~u_i~ \left[\matrix{v\cr v\cr}\right]$ &24 \tabroom \\
VII$_i$ &$g_i~
          \left[0~\matrix{0&0\cr 0&0\cr}\right]
          \left[1~\matrix{1&0\cr 1&0\cr}\right]
          \left[1~\matrix{1&0\cr 1&0\cr}\right]
          \left[1~\matrix{1&0\cr 1&0\cr}\right]
          ~u_i~ \left[\matrix{v\cr v\cr}\right]$ &8 \tabroom \\ [2ex]
\hline
\end{tabular}
\end{center}
\end{scriptsize}
\centerline{ {\bf Table 5:}~{\it Generations of the (80,0) SO(10)
                                 daughter of $3^{\otimes 5}$.}}
\par
The {\bf 10} of SO(10) can be counted by enumerating the singlet part
of the decomposition
${\bf 10} = {\bf 1}_{-2} \oplus {\bf 8}^{s}_0 \oplus {\bf 1}_2$
of the vectors of SO(10) with respect to the maximal subgroup
SO(8)$\times $U(1) which, using abbreviations
\beq
v_0 = \left[2\matrix{-2&0\cr -4&-2\cr}\right],~~~
v_1 = \left[2\matrix{3&1\cr -4&-2\cr}\right]
\lleq{vec-abbr}
for the relevant primary fields in the first factor and
$u_i^+$ for the charge conjugate of $u_i$, leads to the list of
vectors contained in Table 6.
\pano
\begin{scriptsize}
\begin{center}
\begin{tabular}{| l r r|}
\hline
Type   &Field      &Number \tabroom \\
\hline
{\rm A}&$\left[0~\matrix{0&0\cr -2&-2\cr}\right]
         \left[1~\matrix{-1&0\cr -3&-2\cr}\right]
      \left[3~\matrix{-3&0\cr -5&-2\cr}\right]
       \left[3~\matrix{-3&0\cr -5&-2\cr}\right]
      \left[3~\matrix{-3&0\cr -5&-2\cr}\right]
       ~u_0~\left[\matrix{0\cr v\cr}\right]$    &4~  \tabroom \\
{\rm B}&$ \left[0~\matrix{0&0\cr -2&-2\cr}\right]
        \left[2~\matrix{-2&0\cr -4&-2\cr}\right]
      \left[2~\matrix{-2&0\cr -4&-2\cr}\right]
       \left[3~\matrix{-3&0\cr -5&-2\cr}\right]
      \left[3~\matrix{-3&0\cr -5&-2\cr}\right]
      ~u_0~\left[\matrix{0\cr v\cr}\right]$    &6~    \tabroom \\
{\rm C}$_i$&$v_i~
        \left[0~\matrix{0&0\cr -2&-2\cr}\right]
     \left[2~\matrix{-2&0\cr -4&-2\cr}\right]
      \left[3~\matrix{-3&0\cr -5&-2\cr}\right]
     \left[3~\matrix{-3&0\cr -5&-2\cr}\right]
      ~u_i^+~ \left[\matrix{0\cr v\cr}\right]$ &24~    \tabroom \\
{\rm D}$_i$&$v_i~
     \left[1~\matrix{-1&0\cr -3&-2\cr}\right]
 \left[1~\matrix{-1&0\cr -3&-2\cr}\right]
    \left[3~\matrix{-3&0\cr -5&-2\cr}\right]
     \left[3~\matrix{-3& 0\cr -5&-2\cr}\right]
     ~u_i^+~  \left[\matrix{0\cr v\cr}\right]$ &12~   \tabroom \\
{\rm E}$_i$&$v_i~
      \left[1~\matrix{-1&0\cr -3&-2\cr}\right]
      \left[2~\matrix{-2&0\cr -4&-2\cr}\right]
      \left[2~\matrix{-2&0\cr -4&-2\cr}\right]
      \left[3~\matrix{-3&0\cr -5&-2\cr}\right]
      ~u_i^+~ \left[\matrix{0\cr v\cr}\right]$ &24~   \tabroom \\
{\rm F}$_i$ &$v_i~
       \left[2~\matrix{-2&0\cr -4&-2\cr}\right]
       \left[2~\matrix{-2&0\cr -4&-2\cr}\right]
      \left[2~\matrix{-2&0\cr -4&-2\cr}\right]
      \left[2~\matrix{-2&0\cr -4&-2\cr}\right]
      ~u_i^+~  \left[\matrix{0\cr v\cr}\right]$ &2~ \tabroom \\
{\rm G}$_1$ &$ \left[2~\matrix{4&2\cr -4&-2\cr}\right]
               \left[2~\matrix{4&2\cr -4&-2\cr}\right]
               \left[2~\matrix{4&2\cr -4&-2\cr}\right]
               \left[2~\matrix{4&2\cr -4&-2\cr}\right]
               \left[2~\matrix{4&2\cr -4&-2\cr}\right]
      \left[\matrix{2\cr 0\cr}\right] \left[\matrix{0\cr v\cr}\right]$
      &1~ \tabroom \\
{\rm G}$_2$&$\left[2~\matrix{-1&1\cr -4&-2\cr}\right]
             \left[2~\matrix{4&2\cr -4&-2\cr}\right]
               \left[2~\matrix{4&2\cr -4&-2\cr}\right]
               \left[2~\matrix{4&2\cr -4&-2\cr}\right]
               \left[2~\matrix{4&2\cr -4&-2\cr}\right]
            ~u_1^+~ \left[\matrix{0\cr v\cr}\right]$ &1~ \tabroom \\ [2ex]
\hline
\end{tabular}
\end{center}
\end{scriptsize}
\centerline{{\bf Table 6:}~{\it Vectors of the SO(10) (80,0) daughter of
             $3^{\otimes 5}$.}}
With the explicit form of the massless spectrum at our disposal we
can now explore this model in greater depth by computing the Yukawa
couplings. This is what we will turn to in the next Subsection.
\vskip -.1truein
\subsection{The exact Yukawa couplings}
Suppose we wish to compute the Yukawa couplings
\beq
<{\bf 10} \cdot {\bf 16}\cdot {\bf 16}>.
\lleq{gencoup}
Then we have to specify the vertex operators and the picture in which
they live. To get the ghostnumber $-2$ of string theory tree level
correlation functions \cite{fms86} one might consider vertex operators
\beq
<V_{-1}^{\bf 10}~V_{-1/2}^{\bf 16}~V_{-1/2}^{\bf 16}>.
\eeq
Using the decompositions
 \beq
{\bf 10} = {\bf 1}_{-2} \oplus {\bf 8}^{s}_0 \oplus {\bf 1}_2,~~~~
 {\bf 16} = {\bf 8}^v_{-1} \oplus {\bf 8}^c_{1}
\eeq
of the SO(10) representations with respect to the maximal subgroup
SO(8) $\times $ U(1) $\subset$ SO(10), the couplings
$<{\bf 10}\cdot {\bf 16}\cdot {\bf 16}>$ decompose into
\beq
\left< \left(\matrix{{\bf 1}_{-2}\cr {\bf 8}_0^{s}\cr {\bf 1}_2\cr}\right)
       \left(\matrix{{\bf 8}_{-1}^v\cr {\bf 8}_1^c\cr}\right)
       \left(\matrix{{\bf 8}_{-1}^v\cr {\bf 8}_1^c\cr}\right) \right>
= <{\bf 1}_2\cdot {\bf 8}_{-1}^v \cdot {\bf 8}_{-1}^v>
= <{\bf 8}_0^s \cdot {\bf 8}_1^c \cdot {\bf 8}_{-1}^v>.
\eeq
The sum over all right moving charges also has to vanish, a condition
which is not satisfied for the fields listed in Tables 5 and 6.
Using the supersymmetry charge operator $\bSi_{c=9}$ (\ref{right-susy})
of the Gepner parent model one obtains the desired form
\beq
V_{1/2}^a = \bSi~V_{-1}^a
\eeq
with which the Yukawa couplings can be written as
\beq
<V_{-1}^{\bf 10}~V_{-1/2}^{\bf 16}~V_{-1/2}^{\bf 16}>
= <\bSi^2 ({\bf 1}_2)_{-1} ~({\bf 8}^v_{-1})_{-1} ~({\bf 8}^v_{-1})_{-1}>.
\lleq{gen-coup}
\par
The first obstacle any correlation function has to overcome, of course, is
charge conservation, according to which the U(1) charges must cancel
in every factor. Now, according to (\ref{gen-coup}), every coupling of the
type (\ref{gencoup}) contains per construction the operator $\bSi^2$.
We consider first the vectors. Starting with the vector A of Table 6
we need to check which pair of generations we can multiply to get
something nonvanishing. Computing the charges of $\left(\bSi^2~A\right)$
results in
\beq
\left(\matrix{Q\cr \bQ \cr}\right)(\bSi^2~{\rm A}) =
\left(\matrix{0\cr 0\cr}\right)\otimes\left(\matrix{1/5\cr1/5\cr}\right)
\otimes \left(\matrix{3/5\cr 3/5 \cr}\right) \otimes
     \left(\matrix{3/5\cr 3/5 \cr}\right)
\otimes \left(\matrix{3/5\cr 3/5 \cr}\right),
\eeq
hence pairs of generations are needed whose product leads to the negative
of this charge array. Looking back at the generations we see that there are
members of the pair families (I, II) on the one hand, and (II,III) and
(III, IV) on the other, which lead to the appropriate charges.
Thus we have the following potentially nonvanishing Yukawa couplings
\beq
<{\rm A}\cdot {\rm I}\cdot {\rm II}>, ~~~
<{\rm A}\cdot {\rm II}\cdot {\rm III}>,~~~
<{\rm A}\cdot {\rm III} \cdot {\rm IV}>
\eeq
for the simple reason that I -- IV are the only generations with charge
zero in the first factor (the only N$=$2 minimal factor affected by the
simple current). Proceeding in the same manner one finds the remaining
potentially nonvanishing couplings.
\par
In order to compute the actual values of these couplings one uses the
fact that a primary field in the N$=$2 minimal superconformal theory
at level $k$, is just a product of a parafermionic field and a
scalar field $\vphi(z)$
\beq
\Phi^{l,q,s}_{\bl,\bq,\bs}(z,\bz) = \phi^{l,q-s}_{\bl,\bq-\bs}(z,\bz)
     e^{i(\a_{qs} \vphi (z) + \a_{\bq \bs} \vphi'(\bz))},
\eeq
where
\beq
\a_{qs} = \frac{-q +\frac{s}{2}(k+2)}{\sqrt{k(k+2)}}.
\eeq
The parafermionic field $\phi^{l,m}$ in turn can be expressed in terms
of SU(2) primary fields and a further scalar field $\tvphi$ as
\beq
\phi^{l,m} = G^{\frac{l}{2},\frac{m}{2}} ~e^{-\frac{m}{2\sqrt{2}}\tvphi}.
\eeq
Thus the only nontrivial correlation functions one has to know in order
to compute the Yukawa couplings of the superconformal model are
the three--point functions of the SU(2) theory. These have been
obtained by Zamolodchikov and Fateev \cite{zf86}, who found
\beq
<G^{j_1,m_1}_{j_1, \bm_1} G^{j_1,m_1}_{j_1, \bm_1}
      G^{j_1,m_1}_{j_1, \bm_1}>_{{\rm SU}_k}
= \frac{\left(\matrix{j_1&j_2&j_3\cr m_1&m_2&m_3\cr}\right)
        \left(\matrix{j_1&j_2&j_3\cr \bm_1&\bm_2&\bm_3\cr}\right)
        \rho_k(j_1,j_2,j_3)}
       {f(z_{12}\bz_{12}, z_{13}\bz_{13},z_{23}\bz_{23})},
\eeq
where {\scriptsize $\left(\matrix{j_1&j_2&j_3\cr m_1&m_2&m_3\cr}\right)$}
are the Wigner 3$j$--symbols
\bea
& & \left(\matrix{j_1&j_2&j_3\cr m_1&m_2&m_3\cr}\right)
 =\sqrt{\frac{(j_1+j_2-j_3)!(j_1-j_2+j_3)!(-j_1+j_2+j_3)!}
   {(j_1+j_2+j_3+1)!}\prod_{i=1}^3(j_i+m_i)~(j_i-m_i)!}\times\nn\\
& & \times\sum_{z\in \ZZ}
   \frac{(-1)^{z+j_1-j_2-m_3}}{z!(j_1+j_2-j_3-z)!(j_1-m_1-z)!(j_2+m_2-z)!
             (j_3-j_2+m_1+z)!(j_3-j_1-m_2+z)!} \nn  \\
\eea
and
\beq
\rho_k(j_1,j_2,j_3) = F_k(j_1,j_2,j_3)
  \sqrt{\frac{\G \left(\frac{k+3}{k+2}\right)}
   {\G \left(\frac{k+1}{k+2}\right)}
   \prod_{r=1}^3 (2j_r +1)\frac{\G \left(1-\frac{2j_r+ 1}{k+2}\right)}
   {\G \left(1+\frac{2j_r+ 1}{k+2}\right)}},
\eeq
where
\beq
F_k(j_1,j_2,j_3) =
       \frac{\pi_k(j_1+j_2+j_1+1)~\pi_k(j_1+j_2-j_3)~\pi_k(j_2+j_3-j_1)
           \pi_k(j_3+j_1-j_2)} {\pi_k(2j_1)~ \pi_k(2j_2)~ \pi_k(2j_3)}
\eeq
with, finally,
\beq
\pi_k(j) = \prod_{r=1}^j \frac{\G \left(1+\frac{r}{k+2}\right)}
                              {\G \left(1-\frac{r}{k+2}\right)}.
\eeq
\par
Plugging in these ingredients we can compute the necessary Yukawa
couplings of the individual minimal theories. The relevant couplings
for the level $k=3$ theory are
\bea
&
\begin{tabular}{l l}
$\left<\left[1~\matrix{-1&0\cr -1&0\cr}\right]
       \left[1~\matrix{1&0\cr 1&0\cr}\right]
       \left[0~\matrix{0&0\cr 0&0\cr}\right]\right> = 1$
  &~~~~$\left<\left[2~\matrix{-2&0\cr -2&0\cr}\right]
       \left[2~\matrix{2&0\cr 2&0\cr}\right]
       \left[0~\matrix{0&0\cr 0&0\cr}\right]\right> = 1$ \tabroom \\
$\left<\left[3~\matrix{-3&0\cr -3&0\cr}\right]
       \left[3~\matrix{3&0\cr 3&0\cr}\right]
       \left[0~\matrix{0&0\cr 0&0\cr}\right]\right> = 1$
   &~~~~$\left<\left[~3\matrix{-3&0\cr -3&0\cr}\right]
       \left[2~\matrix{2&0\cr 2&0\cr}\right]
       \left[1~\matrix{1&0\cr 1&0\cr}\right]\right> = 1$  \tabroom \\
$\left<\left[2~\matrix{-1&0\cr -1&0\cr}\right]
       \left[1~\matrix{1&0\cr 1&0\cr}\right]
       \left[1~\matrix{1&0\cr 1&0\cr}\right]\right> = \k$
  &~~~~$\left<\left[2~\matrix{3&1\cr -2&0\cr}\right]
        \left[2~\matrix{-3&-1\cr 2&0\cr}\right]
        \left[0~\matrix{0&0\cr 0&0\cr}\right]\right> = 1$, \tabroom \\
\end{tabular}
&
\llea{base-yuk}
where 
\beq
\k = \left(\frac{\G\left(\frac{3}{5}\right)^3 
                 \G\left(\frac{1}{5}\right)} 
                {\G\left(\frac{2}{5}\right)^3 
                 \G\left(\frac{4}{5}\right)}
     \right)^{1/2}.
\eeq
With the couplings for the individual N$=$2 factors at hand
we finally arrive at the Yukawa couplings of the full theory,
the results of which are contained in Table 7.
\pano
\begin{footnotesize}
\begin{center}
\begin{tabular}{|l| l |}
\hline
${\bf 10}$    &Generations  \tabroom \\
\hline
{\rm A}$\cdot$
  &{\rm I}$\cdot ${\rm II}$=1$,~ {\rm II}$\cdot ${\rm III}$=1$,~
                 {\rm III}$\cdot ${\rm IV}$=1$  \tabroom \\
{\rm B}$\cdot$  &{\rm I}$^2=1$,~ {\rm I}$\cdot ${\rm III}$=1$,~
        {\rm II}$^2=\k^2$,~{\rm II}$\cdot ${\rm III}$=\k$,~
        {\rm III}$^2=1$,~{\rm III}$\cdot ${\rm IV}$=\k$,~
        {\rm IV}$^2=\k^2$  \tabroom \\
{\rm C}$_i\cdot$
    &{\rm I}$\cdot ${\rm V}$_i=1$,~ {\rm I}$\cdot ${\rm VI}$_i=1$,~
     {\rm II}$\cdot ${\rm VI}$_i=\k$,~ {\rm III}$\cdot ${\rm VI}$_i=1$,~
       {\rm III}$\cdot ${\rm VII}$_i=\k$  \tabroom \\
{\rm D}$_i\cdot $
      &{\rm I}$\cdot ${\rm VII}$_i=1$,~ {\rm II}$\cdot ${\rm V}$_i=1$,~
      {\rm II}$\cdot ${\rm VI}$_i=1$,~ {\rm III}$\cdot {\rm VII}_i=1$,~
       {\rm IV}$\cdot {\rm VI}_i=1$  \tabroom \\
{\rm E}$_i\cdot $
   &{\rm I}$\cdot ${\rm VI}$_i=1$,~ {\rm II}$\cdot ${\rm VI}$_i=\k$,~
    {\rm II}$\cdot {\rm VII}_i=\k^2$,~ {\rm III}$\cdot {\rm V}_i=1$,~
    {\rm III}$\cdot {\rm VI}_i=1$,~ {\rm III}$\cdot {\rm VII}_i=\k$,~
    {\rm IV}$\cdot {\rm VI}_i=\k$,~ {\rm IV}$\cdot {\rm VII}_i=\k^2$
                   \tabroom \\
{\rm F}$_i\cdot $
    &{\rm III}$\cdot${\rm VI}$_i=\k$,~ {\rm IV}$\cdot ${\rm VII}$_i=\k^3$
                \tabroom \\ [1ex]
\hline
\end{tabular}
\end{center}
\end{footnotesize}
\centerline{{\bf Table 7:}~{\it Yukawa couplings of the generations of the
         SO(10) (80,0) daughter of $3^{\otimes 5}$.}}
\section{Chiral ring and $\si$--model structure of the (80,0) theory}
In this Section we determine the linear $\si$--model whose
fixed point is described by the exactly solvable model
analyzed in depth in the previous Section. This problem naturally splits
into two parts. The first is to determine the chiral ring of the
(0,2) of the theory, defining a monomial algebra, the product structure
of which captures the behaviour of the Yukawa couplings.
In the context of (0,2) theories based either on Calabi--Yau
manifolds or Landau--Ginzburg theories, this would appear to
furnish the appropriate framework in which one eventually should
understand the complete structure of the underlying string model, the
reason being that Calabi--Yau manifolds are algebraic projective
varieties and thus are embedded in a simple space, $\IP_n$, although
by a rather complicated map.
\par
Being able to derive the ring structure from one or more superpotentials,
in order to obtain a complete intersection, simplifies the analysis
considerably, but does restrict the focus to a rather narrow framework,
the confines of which has to be overcome eventually.
In the context of (2,2) compactifications a well known
technique for dealing with rings more general than those originating
from complete intersections is furnished by toric geometry and an
interesting problem for future work would be to formulate (0,2) theories
based on toric varieties. Even toric geometry, however, is a rather
restrictive framework and it would be of great interest to explore
Calabi--Yau manifolds in purely algebraic, ring theoretic, terms.
\par
In the present paper, however, our interest is a different one,
and we focus on the particular exactly solvable theory we have
been analyzing in some detail in the previous Section precisely
because its massless sector is simplified by the fact that it does
not contain antigenerations. Because of this we might hope
that its $\si$--model is described by a simple geometric structure.
We will see below that this is indeed the case and that we are lead
to a rank four vector bundle on a smooth codimension two complete
intersection Calabi--Yau manifold. First we derive the chiral ring.
\vskip -.1truein
\subsection{The (0,2) chiral ring of the (80,0) model}
In order to extract the chiral ring we consider the detailed structure of
the generations. Because the simple current acts only on the first factor
of the tensor product of N$=$2 minimal theories, we expect the chiral
primary fields of the unaffected minimal theories to remain
unchanged. From the structure of the generations of type I -- IV
in Table 5 we indeed see that the basic field which appears
in the last four minimal factors is the chiral primary field
{\scriptsize $\left[1~\matrix{1&0\cr 1&0\cr}\right]$}.
Thus we see that in order to relate these exact states to the complex
coordinates of a ring we should make the identification
\beq
x_i \sim \left[0~\matrix{0&0\cr 0&0\cr}\right] \otimes \cdots \otimes
        \left[1~\matrix{1&0\cr 1&0\cr}\right]
    \otimes \cdots \otimes \left[0~\matrix{0&0\cr 0&0\cr}\right]
    \otimes \left[\matrix{0\cr 0\cr}\right],
\lleq{80x-map}
where the nonzero state is in the $(i+1)st$ factor of the $5$ individual
factors. From the remaining generation families V -- VII of Table 5
we can then read off that the variables
\bea
y_1&\sim &\left[2~\matrix{2&0\cr 2&0\cr}\right]\otimes
    \left[0~\matrix{0&0\cr 0&0\cr}\right] \otimes
     \cdots \otimes \left[0~\matrix{0&0\cr 0&0\cr}\right]
      \otimes \left[\matrix{0\cr 0\cr}\right] \nn \\
y_2&\sim & \left[2~\matrix{-3&-1\cr 2&0\cr}\right] \otimes
     \left[0~\matrix{0&0\cr 0&0\cr}\right] \otimes
     \cdots \otimes \left[0~\matrix{0&0\cr 0&0\cr}\right]
      \otimes \left[\matrix{-1\cr 0\cr}\right]
\llea{80y-map}
must have degree ${\rm deg}(y_i) = 2$, in agreement with the
anomalous weights of the states. It is important to note that for
the `twisted' state $y_2$ the contribution of the U(1)$_2$ is
crucial in order for these fields to be chiral primary fields.
Thus we have two types of coordinates $x_i,~i=1,2,3,4$, coming from
the inert minimal factors and $y_l,~l=1,2$, generated by the
simple current $\Upsilon$.
Using the relations (\ref{80x-map}) and (\ref{80y-map}) maps the
generations into the monomial representations listed in Table 8.
\begin{center}
\begin{tabular}{|l c c|}
\hline
Family    &~~~~  Monomial Representative     &Degeneracy   \tabroom \\
\hline
I   &~~~~  $x_i^2x_j^3$       &$ 12$  \tabroom  \\
II  &~~~~  $x_i x_j x_k^3$    &$ 12$ \tabroom  \\
III &~~~~  $x_i x_j^2 x_k^2$  &$ 12$ \tabroom  \\
IV  &~~~~  $x_ix_jx_kx_l^2$   &$ 4$ \tabroom \\
V$_l$   &~~~~  $x_i^3y_l$         &$ 8$ \tabroom \\
VI$_l$   &~~~~  $x_ix_j^2y_l$      &$ 24$ \tabroom \\
VII$_l$  &~~~~  $x_ix_jx_k y_l$    &$ 8 $ \tabroom \\ [1ex]
\hline
\end{tabular}
\end{center}
\centerline{{\bf Table 8:}~{\it Monomial representation of the generations
                  of the (80,0) model.}}
The chiral ring $\cR = \frac{\IC[x_i,y_j]}{\cI}$ determined by these states
is generated from free ring
\beq
\IC[x_1,x_2,x_3,x_4,y_1,y_2]
\lleq{80free}
by considering equivalence relations defined via the ideal
\beq
\cI [x_1^4, x_2^4, x_3^4, x_4^4, y_1^2, y_2^2, y_1y_2].
\lleq{80ideal}
\par
The question then arises whether this ideal can be derived from
a stable bundle over a Calabi--Yau threefold.
We will show in the next Subsection that this is indeed possible.
First, however, we complete our discussion of the spectrum by
considering the vectors in H$^1(M,\L^2 V)$, which can be identified with
monomials of degree ten. There are a total of 72 of such elements, the
explicit form of which is contained in Table 9.
\begin{center}
\begin{tabular}{|l c c |}
\hline
Family   &~~~~Monomial Representative   &Degeneracy  \tabroom \\
\hline
A        &~~~~$x_ix_j^3x_k^3x_l^3$      &$4$\tab2\\
B        &~~~~$x_i^2x_j^2x_k^3x_l^3$    &$6$\tab2\\
C$_m$        &~~~~$y_mx_i^2x_j^3x_k^3$      &$24$\tab2\\
D$_m$        &~~~~$y_mx_ix_jx_k^3x_l^3$     &$12$\tab2\\
E$_m$        &~~~~$y_mx_ix_j^2x_k^2x_l^3$   &$24$\tab2\\
F$_m$        &~~~~$y_mx_i^2x_j^2x_k^2x_l^2$ &$2$\tab2\\ [1ex]
\hline
\end{tabular}
\end{center}
\centerline{{\bf Table 9:}~{\it Monomial representation of the vectors
                  of the (80,0) model.}}
The precise relation of these monomial forms to the fields of
the exactly solvable theory is indicated by the types A--F, referring
back to Table 6. The two vectors G$_1$, G$_2$ of Table 6
do not have a monomial representation. As we have seen in the
preceding Section they also do not lead to nonvanishing
Yukawa couplings. The product structure of the ring $\cR$ is then
given by the nonvanishing relations described in Table 10.
\begin{center}
\begin{tabular}{|l l |}
\hline
Vector   &Generations   \tabroom \\
\hline
{\rm A}:  &~~~${\rm I}\cdot {\rm II},~~{\rm II}\cdot {\rm III},~~
             {\rm III}\cdot {\rm IV}$ \tabroom  \\
{\rm B}:  &~~~${\rm I}^2,~~{\rm I}\cdot {\rm III},~~{\rm II}^2,~~
           {\rm II}\cdot{\rm III},~~{\rm III}^2,~~{\rm III}\cdot{\rm IV},~~
           {\rm IV}^2$  \tabroom  \\
{\rm C}$_l$:  &~~~${\rm I}\cdot {\rm V}_l,~~{\rm I}\cdot {\rm VI}_l,~~
             {\rm II}\cdot {\rm VI}_l,~~{\rm III}\cdot {\rm VI}_l,~~
             {\rm III}\cdot {\rm VII}_l $  \tabroom  \\
{\rm D}$_l$:&~~~${\rm I}\cdot {\rm VII}_l,~~{\rm II}\cdot {\rm V}_l,~~
             {\rm II}\cdot {\rm VI}_l,~~{\rm III}\cdot {\rm VII}_l,~~
             {\rm IV} \cdot {\rm VI}_l $  \tabroom  \\
{\rm E}$_l$:&~~~${\rm I}\cdot {\rm VI}_l,~~{\rm II} \cdot {\rm VI}_l,~~
             {\rm II} \cdot {\rm VII}_l,~~{\rm III} \cdot {\rm V}_l,~~
             {\rm III} \cdot {\rm VI}_l,~~{\rm III} \cdot {\rm VII}_l,~~
       {\rm IV} \cdot {\rm VI}_l,~~{\rm IV} \cdot {\rm VII}_l $\tabroom\\
{\rm F}$_l$:&~~~${\rm III} \cdot {\rm VI}_l,~~
       {\rm IV} \cdot {\rm VII}_l $\tabroom\\ [1ex]
\hline
\end{tabular}
\end{center}
\centerline{{\bf Table 10:}~{\it Product structure of the ring of the
         (80,0) model.}}
Comparing the product structure of the chiral ring derived in Table 10
with the exactly solvable Yukawas of Table 7 shows
that the simple renormalization
\bea
& & {\rm II} \lra \k {\rm II},~~
    {\rm IV} \lra \k {\rm IV},~~
    {\rm VII}_l \lra \k {\rm VII}_l, \nn\\
& & {\rm A} \lra \k^{-1}~{\rm A},~~
    {\rm D}_l \lra \k^{-1}~{\rm D}_l,~~
    {\rm F}_l \lra \k {\rm F}_l
\llea{10-trafo}
transforms one set of couplings into the other.
It should be noted that in contradistinction to the situation for
the theories discussed in the (2,2) context \cite{g88b,ss88,glr89,rs90}
the transformation is not fixed uniquely but is determined only up
to a two parameter families of rescalings. If, e.g., we define
A $\lra \k^a$ A and VII $\lra \k^x$ VII, then all the remaining
normalizations are determined by the exponents $a$ and $x$.
In (\ref{10-trafo}) we have set $a=-1$ and $x=1$ which seemed to be the
simplest choice because it minimizes the number of renormalizations.
The fact that the map between the Yukawa couplings is not determined
uniquely in the present context will turn out to be a
generic feature even though the degree of nonuniqueness depends
on the specific models, in particular their gauge groups. We
therefore find that knowledge of the underlying exactly solvable
theory is a slightly less powerful tool than it is in the context
of (2,2) theories.
\vskip -.1truein
\subsection{The (0,2) linear $\si$--model}
At this point we have shown that there exists a chiral ring whose product
structure is isomorphic to the Yukawa coupling structure of an exactly
solvable theory. It is then natural to ask whether this ideal can be
derived from the superpotential of a (0,2) $\si$--model. We will show that
indeed it can. Recall \cite{ew93,dk94}, that the essential structure of a
(0,2) linear $\si$--model is encoded in the superpotentials
$(W_r(\Phi_i), F_a^l(\Phi_i)))$, where $W_r(\Phi_i)$ are polynomials of
degree $d_r$ which define the base space $M$ of the vector bundle
$V \lra M$ associated to the left--moving gauge fermions and the
$F^l_a(\Phi_i)$ are polynomials of degree ${\rm deg}F_a^l = m_l - n_a$
which define the global structure of the bundle $V$.
The appropriate constraints are imposed by introducing Lagrange multipliers
$\Si^r $ which are Fermi superfields with charge $(-d_r)$, as well as
Lagrange multipliers $\L^a$ which are Fermi superfields with charges
$n_a$. Finally, one introduces chiral superfields $\Theta_l$ with charges
$m_l$ such that $\sum_l m_l=\sum_a n_a$. The charges assigned to these
fields read as follows
\bea
& &Q(\Phi_i) = k_i,~~i=1,...,N_i;
 ~~~~~~~~~~~~~~~~Q(\Si^r) = -d_l,~~~r=1,...,N_r\nn \\
& &Q(\L^a)  = n_a,~~a=1,...,N_a=r+N_l;~~ Q(\Theta_l) = -m_l,~~~l=1,...,N_l.
\eea
The action which summarizes the structure of the total bundle is then
given by
\beq
\cA = \int d^2z d\th \left[\Si^r W_r(\Phi_i)
       +\Theta_l \L^a F^l_a(\Phi_i)\right].
\eeq
The first term leads to constraints on the $\Phi_i$ to the effect
that they take values on the hypersurface $W_r=0$,
whereas the second term ensures that the gauge fermions $\l^a$
(i.e.\ the lowest components of the $\L^a$) are sections of the bundle $V$.
\par
The structure of the bundle $V$ can be summarized concisely by the
short exact sequence
\beq
0 \lra V \lra \bigoplus_{a=1}^{r+N_l} \cO(n_a) \mathop{\lra }^F
     \bigoplus_{l=1}^{N_l} \cO(m_l) \lra 0
\eeq
which allows to compute the Chern class
\beq
c(V) = \frac{ c\left( \bigoplus_{a=1}^{r+N_l} \cO(n_a)\right)}
            { c\left( \bigoplus_{l=1}^{N_l} \cO(m_l)\right)}
\eeq
which, with
\beq
c\left( \bigoplus_{a=1}^{r+N_l} \cO(n_a)\right) =
 \prod_a c(\cO(n_a)) = \prod_a (1+c_1( \cO(n_a)) = \prod_a (1+n_a h)
\eeq
and
\beq
c(\bigoplus_{l=1}^{N_l} \cO(m_l)) = \prod_l c(\cO(m_l)) =\prod_l (1+m_lh),
\eeq
leads to
\beq
c(V) = \frac{\prod_a (1+n_a h)}{\prod_l (1+m_lh)}.
\eeq
Expanding this expression leads to the individual Chern classes
\bea
c_1(V) &=& \left[\sum_a n_a - \sum_l m_l\right]h \nn \\
c_2(V) &=& \frac{1}{2}\left[\sum_l m_l^2-\sum_a n_a^2\right]h^2 \nn \\
c_3(V) &=& -\frac{1}{3} \left[\sum_l m_l^3 - \sum_a n_a^3\right]h^3.
\eea
Recalling that for a complete intersection of the form
$\IP_{(k_1,...,k_{N_i})}\left[d_1~\cdots d_{N_r}\right] $
the second Chern class is given by
$c_2(TM)=\frac{1}{2}\left[\sum_r d_r^2-\sum_i k_i^2\right]$,
the anomaly matching condition can be written as
\beq
\sum_l m_l^2 - \sum_a n_a^2 = \sum_r d_r^2-\sum_i k_i^2.
\eeq
\par
Now, the structure of the ring $\cR$ given by (\ref{80free}) and
(\ref{80ideal}) tells us that we should look for potentials $(W_r, F_a^l)$
defining a bundle $V$ of rank four which are of degree four. As derived in
the previous Subsection the base space is spanned by four coordinates
$x_i,~i=1,..,4$ of unit weight and two coordinates $y_l,~l=1,2$ of
weight two. Since we wish to consider a 3--fold it follows that we need
to impose two polynomials $W_1,W_2$ of degree four, i.e.\ the bundle $V$
will live on the base configuration $M= \IP_{(1,1,1,1,2,2)}[4~4]$. This
manifold is indeed a Calabi--Yau manifold, i.e.\ $c_1(M)=0$, the Chern
classes of which are $c_2(M)=10h^2$ and $c_3(M)=-36h^3$, where $h$ is
the pullback of the generator of H$^2(M)$.
\par
The simplest way to define a rank four bundle with potentials
$F_a$ of degree four is via the short sequence
\beq
0 \lra V \lra \bigoplus_{a=1}^{5} \cO(n_a=1) \mathop{\lra }^F
     \cO(5) \lra 0,
\eeq
the resulting bundle of which we denote by $V_{(1,1,1,1,1;5)}$,
adopting the convention of \cite{sk95}. The second Chern class of this
bundle is equal to the second Chern class of the manifold and hence the
anomalies cancel. Furthermore $c_1(V)=0$ and therefore the bundle
\beq
V_{(1,1,1,1,1;5)} \lra \IP_{(1,1,1,1,2,2)}[4~4]
\eeq
describes the Calabi--Yau phase of the linear $\si$--model defined
by the chiral primary fields with the charges $Q(\Phi_i)=1,~i=1,..,4$,
$Q(\Phi_i)=2,~i=5,6$ and $Q(\L^a)=1,~a=1,..,5$. With
$c_3(V)=-160h^3$ one finds for the Euler number
\beq
\chi(V_{(1,1,1,1,1;5)}) = -160
\eeq
and because the manifold is smooth we expect that this theory
has no antigenerations. It follows from the structure of the chiral
ring that this is indeed the case and therefore this $\si$--model leads
to 80 generations, in agreement with the exact model discussed in
Section 3.
\par
The problem then boils down to finding a choice for the defining
polynomials of the base and the defining data for the bundle
$V_{(1,1,1,1,1;5)}$ which leads to the ideal (\ref{80ideal}).
First one has to pick the defining data of the manifold,
i.e.\ the polynomials $W_r$. A transverse choice is given by
\bea
W_1 &=& \sum_i x_i^4 + \sum_l y_l^2 \nn \\
W_2 &=& \sum_i ix_i^4 + \sum_l l y_l^2 .
\eea
Next we need to pick the bundle data, i.e.\ the $F_a$s, which we do as
\beq
(F_a) = (x_1^4, x_2^4, x_3^4, x_4^4, y_1y_2).
\eeq
One easily checks that the ideal generated by $(W_r, F_a)$ then is given by
\beq
\cI [x_1^4, x_2^4, x_3^4, x_4^4, y_1^2, y_2^2, y_1y_2].
\eeq
This is precisely the ideal we have derived previously from the structure
of the generations in the exactly solvable theory.
\par
Having succeeded in identifying the (0,2) Calabi--Yau manifold, the
conformal fixed point of which is described by the exactly solvable (80,0)
theory, we realize that the product structure we have computed in
Section 4.1 for the chiral ring derived from the exact model
is nothing but the product structure
of the cohomology ring H$^p(M,\L^q V)$. In particular the Yukawa
couplings are determined by the product
\beq
{\rm H}^1(M,V) \otimes {\rm H}^1(M,V) \lra {\rm H}^1(M,\L^2 V),
\eeq
where we have used the isomorphism
H$^p(M,\L^q V) =$ H$^{D-p}(M,\L^{r-q} V)$ for a bundle $V$ of rank $r$ on
a $D$--dimensional variety.
\par
The linear $\si$--model that we have thus derived from the exactly
solvable model of Section 3 has been shown \cite{dk94}
to contain both a (0,2) Calabi--Yau as well as a Landau--Ginzburg phase.
We therefore have provided for the SO(10)--(80,0) (0,2)--theory
 all the ingredients familiar from the context of (2,2) exactly solvable
Calabi--Yau $\si$--models. In particular we have established the
existence of exactly solvable theories describing the conformal
fixed points of (0,2) Calabi--Yau linear $\si$--models and have
described its precise nature.
\section{An SU(5) (64,0) (0,2)--model derived from $3^{\otimes 5}$}
\vskip -.1truein
\subsection{The exactly solvable theory}
The exact model is derived from the parent tensor theory
$3^{\otimes 5}$ by inserting the two simple currents
\bea
\Upsilon_1 &=& [3~0~-1]\otimes [0~0~0] \otimes [0~0~0]^{\otimes 3}
             \otimes [1] \otimes [0] \otimes [0] \nn \\
\Upsilon_2 &=& [3~0~-1]\otimes [3~0~-1] \otimes [0~0~0]^{\otimes 3}
             \otimes [1] \otimes [1] \otimes [0].
\eea
In this model the E$_6$ of the Gepner model is broken down to SU(5)
and hence the generation/antigeneration structure is determined by the
${\bf 10}$ and ${\bf {\overline {10}}}$ representations of SU(5).
The massless spectrum of this model contains (among others) the
modes of Table 11.
\pano
\begin{small}
\begin{center}
\begin{tabular}{|l r r r r r|}
\hline
Representation &${\bf 0}$ &${\bf 10}$ &${\bf {\overline {10}}}$ &${\bf 5}$
                                       &${\bf {\bar 5}}$ \tabroom  \\
\hline
Spin 0          &338        &64         &0        &55      &119\tabroom \\
Spin 1          &10         &0          &0        &0       &0  \tabroom \\
\hline
\end{tabular}
\end{center}
\end{small}
\centerline{{\bf Table 11:}~{\it Massless spectrum of the
         (64,0) SU(5) daughter of the $3^{\otimes 5}$ model.}}
Now, recall that the ${\bf 10}$ decomposes with respect to the maximal
subgroup SO(6)$\times $U(1) as
\beq
{\bf 10} = {\bf 6}_{-1} \oplus {\bf 4}_{3/2}.
\eeq
The generations that result are enumerated in Table 12, where we again
use the abbreviations $g_i$ of (\ref{gen-abbr}) for the two states
appearing in those two minimal N$=$2 factors at level three that are
affected by the simple current as well as the notation (\ref{u1-abbr})
for the U(1) factors.
\pano
\begin{scriptsize}
\begin{center}
\begin{tabular}{| l r r|}
\hline
Type   &Field      &Number \tabroom \\
\hline
I  &$\left[0~\matrix{0&0\cr 0&0\cr}\right]
     \left[0~\matrix{0&0\cr 0&0\cr}\right]
     \left[0~\matrix{0&0\cr 0&0\cr}\right]
     \left[2~\matrix{2&0\cr 2&0\cr}\right]
    \left[3~\matrix{3&0\cr 3&0\cr}\right]
    ~u_0~u_0~\left[\matrix{v\cr v\cr}\right]$ &6 \tabroom \\
II & $\left[0~\matrix{0&0\cr 0&0\cr}\right]
      \left[0~\matrix{0&0\cr 0&0\cr}\right]
      \left[1~\matrix{1&0\cr 1&0\cr}\right]
      \left[1~\matrix{1&0\cr 1&0\cr}\right]
      \left[3~\matrix{3&0\cr 3&0\cr}\right]
     ~u_0~u_0~\left[\matrix{v\cr v\cr}\right]$ &3 \tabroom \\
III & $\left[0~\matrix{0&0\cr 0&0\cr}\right]
     \left[0~\matrix{0&0\cr 0&0\cr}\right]
    \left[1~\matrix{1&0\cr 1&0\cr}\right]
      \left[2~\matrix{2&0\cr 2&0\cr}\right]
    \left[2~\matrix{2&0\cr 2&0\cr}\right]
     ~u_0~u_0~\left[\matrix{v\cr v\cr}\right]$ &3 \tabroom \\
IV$_i$ &$g_i\left[0~\matrix{0&0\cr 0&0\cr}\right]
    \left[0~\matrix{0&0\cr 0&0\cr}\right]
      \left[0~\matrix{0&0\cr 0&0\cr}\right]
        \left[3~\matrix{3&0\cr 3&0\cr}\right]
      ~u_i~u_0~ \left[\matrix{v\cr v\cr}\right]$ &6 \tabroom \\
V$_i$ &$g_i
     \left[0~\matrix{0&0\cr 0&0\cr}\right]
     \left[0~\matrix{0&0\cr 0&0\cr}\right]
     \left[1~\matrix{1&0\cr 1&0\cr}\right]
         \left[2~\matrix{2&0\cr 2&0\cr}\right]
       ~u_i~u_0\left[\matrix{v\cr v\cr}\right]$ &12 \tabroom \\
VI$_i$ &$g_i
          \left[0~\matrix{0&0\cr 0&0\cr}\right]
          \left[1~\matrix{1&0\cr 1&0\cr}\right]
          \left[1~\matrix{1&0\cr 1&0\cr}\right]
          \left[1~\matrix{1&0\cr 1&0\cr}\right]
       ~u_i~u_0~\left[\matrix{v\cr v\cr}\right]$ &2 \tabroom \\
VII$_i$&$\left[0~\matrix{0&0\cr 0&0\cr}\right]
        g_i ~\left[0~\matrix{0&0\cr 0&0\cr}\right]
        \left[0~\matrix{0&0\cr 0&0\cr}\right]
        \left[3~\matrix{3&0\cr 3&0\cr}\right]
       ~u_0~u_i~\left[\matrix{v\cr v\cr}\right]$ &6  \tabroom \\
VIII$_i$ &$ \left[0~\matrix{0&0\cr 0&0\cr}\right]
      g_i~\left[0~\matrix{0&0\cr 0&0\cr}\right]
        \left[1~\matrix{1&0\cr 1&0\cr}\right]
         \left[2~\matrix{2&0\cr 2&0\cr}\right]
       ~u_0~u_i~\left[\matrix{v\cr v\cr}\right]$ &12 \tabroom \\
IX$_i$ &$ \left[0~\matrix{0&0\cr 0&0\cr}\right]
          g_i~\left[1~\matrix{1&0\cr 1&0\cr}\right]
          \left[1~\matrix{1&0\cr 1&0\cr}\right]
          \left[1~\matrix{1&0\cr 1&0\cr}\right]
     ~u_0~u_i~\left[\matrix{v\cr v\cr}\right]$ &2 \tabroom \\
X$_{ij}$ &$ g_i~g_j
         \left[0~\matrix{0&0\cr 0&0\cr}\right]
         \left[0~\matrix{0&0\cr 0&0\cr}\right]
         \left[1~\matrix{1&0\cr 1&0\cr}\right]
        ~u_i~u_j~ \left[\matrix{v\cr v\cr}\right]$ &12 \tabroom \\ [2ex]
\hline
\end{tabular}
\end{center}
\end{scriptsize}
\centerline{{\bf Table 12:}~{\it Generations of the SU(5) (64,0)--daughter
                                  of $3^{\otimes 5}$.}}
The ${\bf 5}$s of the SU(5) decompose with respect to the
maximal subgroup SO(6)$\times $U(1) as
\beq
{\bf 5} = {\bf 4}_{-1/2} \oplus {\bf 1}_{2}.
\eeq
Counting the singlet part and recalling the abbreviations (\ref{vec-abbr})
the 55 representations of the {\bf 5} then take the form presented in
Table 13.
\pano
\begin{scriptsize}
\begin{center}
\begin{tabular}{| l r r |}
\hline
Type   &Field  &Number \tabroom \\
\hline
{\rm A}$_i$
    &$v_i \left[0\matrix{0&0\cr -2&-2\cr}\right]
      \left[2\matrix{-2&0\cr -4&-2\cr}\right]
   \left[3\matrix{-3&0\cr -5&-2\cr}\right]
    \left[3\matrix{-3&0\cr -5&-2\cr}\right]~u_i^+~u_0~
       \left[\matrix{0\cr v\cr}\right]$ &6~ \tabroom \\
{\rm B}$_i$
   &$\left[0\matrix{0&0\cr -2&-2\cr}\right]
      ~v_i~\left[2\matrix{-2&0\cr -4&-2\cr}\right]
     \left[3\matrix{-3&0\cr -5&-2\cr}\right]
      \left[3\matrix{-3&0\cr -5&-2\cr}\right]~u_0~u_i^+~
    \left[\matrix{0\cr v\cr}\right]$ &6~ \tabroom \\
{\rm C}$_{ij}$
 &$v_i~v_j~ \left[0\matrix{0&0\cr -2&-2\cr}\right]
\left[3\matrix{-3&0\cr -5&-2\cr}\right]
    \left[3\matrix{-3&0\cr -5&-2\cr}\right]u_i^+~u_j^+~
      \left[\matrix{0\cr v\cr}\right] $ &12~ \tabroom \\
{\rm D}$_{ij}$
&$v_i~v_j~ \left[3\matrix{-3&0\cr -5&-2\cr}\right]
          \left[2\matrix{-2&0\cr -4&-2\cr}\right]
     \left[1\matrix{-1&0\cr -3&-2\cr}\right]u_i^+~u_j^+~
        \left[\matrix{0\cr v\cr}\right]$ &24~ \tabroom \\
{\rm E}$_{ij}$
&$ v_i~v_j~
  \left[2\matrix{-2&0\cr -4&-2\cr}\right]
     \left[2\matrix{-2&0\cr -4&-2\cr}\right]
    \left[2\matrix{-2&0\cr -4&-2\cr}\right]u_i^+~u_j^+~
        \left[\matrix{0\cr v\cr}\right]$ &4~ \tabroom \\
{\rm F}
&$\left[3~\matrix{-4&-1\cr -5&-2\cr}\right]
   \left[3~\matrix{-4&-1\cr -5&-2\cr}\right]
   \left[1~\matrix{3&2\cr -3&-2\cr}\right]
   \left[1~\matrix{3&2\cr -3&-2\cr}\right]
   \left[2~\matrix{2&0\cr -4&-2\cr}\right]
    ~u_1^+~u_1^+~\left[\matrix{0\cr v\cr}\right]$ &3~ \tabroom \\ [2ex]
\hline
\end{tabular}
\end{center}
\end{scriptsize}
\centerline{{\bf Table 13:}~{\it The {\bf 5}s of the {\rm SU(5)}
                  (64,0) (0,2) model of $3^5$.}}
We now proceed to the couplings.
\vskip -.1truein
\subsection{The exact Yukawa couplings}
Recalling the decomposition ${\bf 10}={\bf 6}_{-1}\oplus{\bf 4}_{3/2}$
one notes that the allowed nonvanishing combination for the Yukawa
couplings can be taken to be of the form
\beq
<{\bf 5} \cdot{\bf 10}\cdot {\bf 10}> ~~= ~~
<{\bf 1}_{2} \cdot {\bf 6}_{-1} \cdot {\bf 6}_{-1} >.
\eeq
Getting the pictures aligned again involves
\begin{scriptsize}
$\bSi =  \left[0~\matrix{0&0\cr 1&1\cr}\right]^{\otimes 5}$,
\end{scriptsize}
and following our discussion of the first example we find that the
only nonvanishing Yukawa couplings involving, say, the field A, are
given by
\beq
<{\rm A}_1\cdot {\rm I}\cdot {\rm IV}_1>,~~~
<{\rm A}_1\cdot {\rm I}\cdot {\rm V}_1>,~~~
<{\rm A}_1\cdot {\rm II}\cdot {\rm V}_1>,~~~
<{\rm A}_1\cdot {\rm III}\cdot {\rm V}_1>,~~~
<{\rm A}_1\cdot {\rm III}\cdot {\rm VI}_1>.
\eeq
For the explicit evaluation of these couplings it again suffices to
consider the basic couplings in the individual N$=$2, $k=3$ minimal
theory collected in (\ref{base-yuk}).
Proceeding in this manner one finds again all possibly nonvanishing
Yukawas. The result is summarized in Table 14.
\pano
\begin{footnotesize}
\begin{center}
\begin{tabular}{|l| l |}
\hline
${\bf 5}$    &Generations  \tabroom \\
\hline
{\rm A}$_i\cdot$
     &{\rm I}$\cdot ${\rm IV}$_i=1$,~~{\rm I}$\cdot ${\rm V}$_i=1$,~~
      {\rm II}$\cdot ${\rm V}$_i=\k$,~~{\rm III}$\cdot ${\rm V}$_i=1$,~~
      {\rm III}$\cdot ${\rm VI}$_i=\k$  \tabroom \\
{\rm B}$_i \cdot$
&{\rm I}$\cdot ${\rm VII}$_i=1$,~~{\rm I}$\cdot ${\rm VIII}$_i=1$,~~
  {\rm II}$\cdot ${\rm VIII}$_i=\k$,~~{\rm III}$\cdot ${\rm VIII}$_i=1$,~~
          {\rm III}$\cdot ${\rm IX}$_i=\k$  \tabroom \\
{\rm C}$_{ij}\cdot $
   &{\rm I}$ \cdot ${\rm X}$_{ij}=1$,~~ {\rm IV}$_i\cdot {\rm VII}_j=1$,~~
              {\rm V}$_i\cdot {\rm VIII}_j=1$  \tabroom \\
{\rm D}$_{ij}\cdot $
  &{\rm I}$ \cdot ${\rm X}$_{ij}=1$,~~{\rm II}$ \cdot ${\rm X}$_{ij}=\k$,~~
  {\rm III}$ \cdot ${\rm X}$_{ij}=1$,~~{\rm IV}$_i \cdot {\rm VIII}_j=1$,~~
  {\rm V}$_i\cdot {\rm VII}_j=1$,~~{\rm V}$_i\cdot {\rm IX}_j=\k$,~~
                {\rm VI}$_i\cdot {\rm VIII}_j=\k$  \tabroom \\
{\rm E}$_{ij}\cdot $
  &{\rm III}$ \cdot {\rm X}_{ij}=\k$,~~ {\rm V}$_i\cdot {\rm VIII}_j=\k$,~~
          {\rm VI}$_i\cdot {\rm IX}_j=\k^3$ \tabroom \\ [1ex]
\hline
\end{tabular}
\end{center}
\end{footnotesize}
\centerline{{\bf Table 14:}~{\it Yukawa couplings of the (64,0)
             SU(5) theory.}}
As indicated in Table 14 there are no couplings involving the modes F$_i$.
\vskip -.1truein
\subsection{The chiral ring of the (64,0) SU(5) model:\ states and
             product structure}
The structure of the chiral ring for the present model can be derived in a
way completely analogous to our discussion of the first example.
Consider again the structure of those generations that do not contain
contributions from the minimal factors affected by the simple current.
These suggest that again we should introduce coordinates $x_i$ related to
the exact state
\beq
x_i \sim \left[0~\matrix{0&0\cr 0&0\cr}\right] \otimes \cdots \otimes
        \left[1~\matrix{1&0\cr 1&0\cr}\right]
    \otimes \cdots \otimes \left[0~\matrix{0&0\cr 0&0\cr}\right]
    \otimes \left[\matrix{0\cr 0\cr}\right]
    \otimes \left[\matrix{0\cr 0\cr}\right],
\eeq
where the nonzero state is in the $(i+2)st$ factor of the $n$ individual
factors. Again we can read off from the structure of the generations that
the weight of the corresponding fields is unity. From the remaining
generations we can then read off that the variables
\bea
y_1&\sim &\left[2~\matrix{2&0\cr 2&0\cr}\right]\otimes
    \left[0~\matrix{0&0\cr 0&0\cr}\right] \otimes
     \cdots \otimes \left[0~\matrix{0&0\cr 0&0\cr}\right]
      \otimes \left[\matrix{0\cr 0\cr}\right]
      \otimes \left[\matrix{0\cr 0\cr}\right] \nn \\
y_2&\sim & \left[2~\matrix{-3&-1\cr 2&0\cr}\right] \otimes
     \left[0~\matrix{0&0\cr 0&0\cr}\right] \otimes
     \cdots \otimes \left[0~\matrix{0&0\cr 0&0\cr}\right]
      \otimes \left[\matrix{-1\cr 0\cr}\right]
      \otimes \left[\matrix{0\cr 0\cr}\right] \nn \\
z_1&\sim &\left[0~\matrix{0&0\cr 0&0\cr}\right] \otimes
      \left[2~\matrix{2&0\cr 2&0\cr}\right]\otimes
     \cdots \otimes \left[0~\matrix{0&0\cr 0&0\cr}\right]
      \otimes \left[\matrix{0\cr 0\cr}\right]
      \otimes \left[\matrix{0\cr 0\cr}\right] \nn \\
z_2&\sim & \left[0~\matrix{0&0\cr 0&0\cr}\right] \otimes
        \left[2~\matrix{-3&-1\cr 2&0\cr}\right] \otimes
     \cdots \otimes \left[0~\matrix{0&0\cr 0&0\cr}\right]
      \otimes \left[\matrix{0\cr 0\cr}\right]
      \otimes \left[\matrix{-1\cr 0\cr}\right]
\eea
must have degree ${\rm deg}(y_i)=2={\rm deg}(z_i)$. With these coordinates
the generations described in Subsection 5.1 are mapped into the monomials
of Table 15.
\pano
\begin{small}
\begin{center}
\begin{tabular}{|l c c|}
\hline
Family  &Monomial Representative  &Degeneracy  \tabroom \\
\hline
I                &~~~~$x_i^2x_j^3$       &$6$\tabroom \\
II               &~~~~$x_i x_j x_k^3$    &$3$\tabroom \\
III              &~~~~$x_i x_j^2 x_k^2$  &$3$\tabroom \\
IV$_l$     &~~~~$y_lx_i^3$         &$6$\tabroom \\
V$_l$      &~~~~$y_lx_ix_j^2$      &$12$\tabroom \\
VI$_l$     &~~~~$y_lx_ix_jx_k$     &$2$\tabroom \\
VII$_m$    &~~~~$z_mx_i^3$         &$6$\tabroom \\
VIII$_m$   &~~~~$z_mx_ix_j^2$      &$12$\tabroom \\
IX$_m$     &~~~~$z_mx_ix_jx_k$     &$2$\tabroom \\
X$_{l,m}$    &~~~~$y_lz_mx_i$        &$12$\tabroom \\ [1ex]
\hline
\end{tabular}
\end{center}
\end{small}
\centerline{{\bf Table 15:}~{\it Monomial representation of the generations
      of the (64,0) model.}}
It follows from the form of the generations that the ideal $\cI$ which
generates the chiral ring $\cR  =  \frac{\IC[x_i,y_j,z_k]}{\cI}$ from the
free ring $\IC[x_i,y_j,z_k]$ is generated by
\bea
x_i^4 &=&0,~~~i=1,2,3 \nn \\
y_ly_{l'} &=&0,~~~l,l'=1,2 \nn \\
z_mz_{m'} &=&0,~~~m,m'=1,2.
\eea
Next we consider the {\bf 5} representation, i.e.\ the monomials of
degree ten, which take the form summarized in Table 16.
\begin{center}
\begin{tabular}{|l c c|}
\hline
Family  &Monomial Representative   &Degeneracy  \tabroom \\
\hline
A$_l$           &~~~~$y_lx_i^2x_j^3x_k^3$    &$6$\tabroom \\
B$_m$           &~~~~$z_mx_i^2x_j^3x_k^3$    &$6$\tabroom \\
C$_{l,m}$       &~~~~$y_lz_mx_i^3x_j^3$      &$12$\tabroom \\
D$_{l,m}$       &~~~~$y_lz_mx_ix_j^2x_k^3$   &$24$\tabroom \\
E$_{l,m}$       &~~~~$y_lz_mx_i^2x_j^2x_k^2$ &$4$\tabroom \\ [1ex]
\hline
\end{tabular}
\end{center}
\centerline{{\bf Table 16:}~{\it Monomial representation of the vectors
                  of the (64,0) model.}}
Thus there are a total of 52 vectors which can be represented by monomials.
It is clear from their structure that for all generations
$\g_i^2 \in \cI$, where $\cI$ is the ideal described above. Thus there
exists no nonvanishing Yukawa `selfcoupling' of any generation.
The Yukawas follow from
\beq
\g_i \cdot \g_j \sim v_{ij},
\eeq
where $v_{ij}$ is any of the {\bf 5}s. With this one finds the
nonvanishing couplings of Table 17.
\begin{center}
\begin{tabular}{|l l|}
\hline
Vector   &Generations  \tabroom \\
\hline
A$_l$:     &~~~${\rm I}\cdot {\rm IV}_l,~~{\rm I}\cdot {\rm V}_l,~~
               {\rm II}\cdot {\rm V}_l,~~{\rm III}\cdot {\rm V}_l,~~
               {\rm III}\cdot {\rm VI}_l$ \tabroom   \\
B$_l$:     &~~~${\rm I}\cdot {\rm VII}_l,~~{\rm I}\cdot {\rm VIII}_l,~~
               {\rm II}\cdot {\rm VIII}_l,~~{\rm III}\cdot {\rm VIII}_l,~~
               {\rm III}\cdot {\rm IX}_l$ \tabroom  \\
C$_{lm}$: &~~~${\rm I}\cdot{\rm X}_{l,m},~~{\rm IV}_l\cdot{\rm VII}_m,~~
               {\rm V}_l \cdot {\rm VIII}_m $ \tabroom \\
D$_{lm}$: &~~~${\rm I}\cdot{\rm X}_{lm},~~{\rm II}
                \cdot{\rm X}_{lm},~~{\rm III}\cdot{\rm X}_{lm},
                ~~{\rm IV}_l\cdot{\rm VIII}_m,~~{\rm V}_l
                \cdot {\rm VII}_m,~~{\rm V}_l\cdot{\rm IX}_m,
                ~~{\rm VI}_l\cdot{\rm VIII}_m $ \tabroom \\
E$_{lm}$:   &~~~${\rm III}\cdot {\rm X}_{lm},~~{\rm V}_l\cdot {\rm VIII}_m,
                 ~~ {\rm VI}_l \cdot {\rm IX}_m$ \tabroom \\ [1ex]
\hline
\end{tabular}
\end{center}
\centerline{{\bf Table 17:}~{\it Product structure of the chiral ring
            of the (64,0) SU(5) theory.}}
Renormalizing the fields as
\beq
{\rm II} \lra \k {\rm II},~~~{\rm VI}_l \lra \k {\rm VI}_l,~~~
{\rm IX}_l \lra \k {\rm IX}_l,~~~{\rm E}_{lm}\lra \k {\rm E}_{lm}
\eeq
then transforms the sigma model couplings into the exactly solvable
couplings. As in the (80,0) model this transformation is not unique,
the difference being that now we have three scales at our disposal.
\vskip -.1truein
\subsection{Remarks concerning the linear $\si$--model}
As in the previous Section we wish to identify superpotentials
$(W_r(\Phi_i), F^l_a(\Phi_i))$ which in the present case define a
rank five bundle over a base space spanned by coordinates
$(x_i, y_l, z_m),~i=1,2,3;~l=1,2;~m=1,2.$ The structure of the chiral
ring suggests that we again consider potentials $W_r$ of degree $d_r=4$.
This does not seem to work however because we have seven coordinates of
weights $k_i$ such that the sum of the weights is eleven. Therefore
it appears that in order to define a Calabi--Yau threefold we have
to introduce three polynomials $W_r$ of degree $d_r$ such that
$\sum d_r=11$. This leads one to introduce a further coordinate of
weight five and to consider the configuration
\beq
\IP_{(1,1,1,2,2,2,2,5)}[4~4~4~4].
\eeq
The virtue of this space is that it has a number of pleasant properties.
First, it satisfies the anomaly condition
\beq
\sum_i k_i = \sum_r d_r,
\eeq
thereby defining a complete intersection Calabi--Yau manifold of
codimension 4. Second, it satisfies a condition familiar from the context
of Landau--Ginzburg theories \cite{ls88}. Namely the total charge of the
theory is the codimension of the corresponding $\si$--model.
Third, it does not have any orbifold singularities. This is an
expected feature if the K\"ahler sector is to be absent for a
(0,2) Calabi--Yau manifold. The disadvantage is that this space does not
allow for a transverse choice of polynomials. It has a hypersurface
singularity at one point. We hope that future insight into the resolution
of this singularity will in fact turn this apparent difficulty into a
virtue which will resolve a second puzzle which this space generates.
\par
As in the theory discussed in Sections 3 and 4 we have been lead
to a ring which is generated by an ideal spanned by elements of
degree four. All that has happened is that the number of generators
has increased, a change which we have already incorporated in our
choice of the base space. It thus would appear natural
to consider a bundle with the same quantum numbers of the gauge
fermions as before. This leads us to
\beq
V_{(1,1,1,1,1;5)} \lra \IP_{(1,1,1,2,2,2,2,5)}[4~4~4~4].
\eeq
Furthermore, the condition involving the second Chern classes
of the gauge vector bundle and the tangent bundle
is met by this choice of a bundle and therefore the
linear $\si$--model defined by this structure satisfies all of
the anomaly conditions. If the bundle would be defined over a smooth
manifold however, it would be of rank four and not of rank five.
We have to leave for future work whether a better understanding of
the resolution does in fact cancel these two difficulties and make
this bundle into a proper (0,2) Calabi--Yau manifold or whether an
altogether different bundle can be found.
\section{The (50,0) SU(3)$\times $SU(2) (0,2)--model derived
       from $3^{\otimes 5}$}
\vskip -.1truein
\subsection{The exact theory}
Consider the quintic tensor model $3^{\otimes 5}$
enhanced with the three simple currents
\bea
\Upsilon_1&=&[3~0~-1]\otimes[0~0~0]\otimes[0~0~0]\otimes[0~0~0]^{\otimes2}
 \otimes[1]\otimes[0]\otimes[0]\otimes[0] \nn \\
\Upsilon_2&=&[3~0~-1]\otimes[3~0~-1]\otimes[0~0~0]\otimes[0~0~0]^{\otimes2}
 \otimes[1]\otimes[1]\otimes[0]\otimes[0] \nn \\
\Upsilon_3&=&[3~0~-1]\otimes[3~0~-1]\otimes[3~0~-1]
 \otimes[0~0~0]^{\otimes2}\otimes[1]\otimes[1]\otimes[1]\otimes[0].
\eea
In this model the E$_6$ of the parent Gepner model is broken to
SU(3)$\times $ SU(2), hence the modes are arranged in representations
of SU(3)$\times $ SU(2) as shown for the massless sector in Table 18.
\pano
\begin{small}
\begin{center}
\begin{tabular}{|l c c c c c c|}
\hline
Representation: &{\bf 0} &{\bf 2} &{\bf 3} &${\bf {\bar 3}}$ &{\bf 6}
                                  &${\bf {\overline {6}}}$  \tabroom \\
\hline
Spin 0:         & 370      &134        &54       &154  &50 &0 \tabroom\\
Spin 1:         & 13       & 0         &0        &0    &0  &0 \tabroom\\
\hline
\end{tabular}
\end{center}
\end{small}
\centerline{{\bf Table 18:}~{\it Massless spectrum of the (50,0)
           SU(3)$\times $SU(2) model.}}
The form of the generations can be found in Table 19.
\pano
\begin{scriptsize}
\begin{center}
\begin{tabular}{| l r r|}
\hline
Type   &Field      &Number \tabroom \\
\hline
{\rm I}&$\left[0~\matrix{0&0\cr 0&0\cr}\right]
         \left[0~\matrix{0&0\cr 0&0\cr}\right]
         \left[0~\matrix{0&0\cr 0&0\cr}\right]
         \left[2~\matrix{2&0\cr 2&0\cr}\right]
         \left[3~\matrix{3&0\cr 3&0\cr}\right]
         ~u_0^3~\left[\matrix{v\cr v\cr} \right]$ &2~ \tabroom \\
{\rm II}$_i$ &$ g_i~ \left[0~\matrix{0&0\cr 0&0\cr}\right]
                \left[0~\matrix{0&0\cr 0&0\cr}\right]
                \left[0~\matrix{0&0\cr 0&0\cr}\right]
                \left[3~\matrix{3&0\cr 3&0\cr}\right]
          ~u_i~u_0^2~\left[\matrix{v\cr v\cr} \right]$ &4~ \tabroom \\
{\rm III}$_i$ &$ g_i~ \left[0~\matrix{0&0\cr 0&0\cr}\right]
                 \left[0~\matrix{0&0\cr 0&0\cr}\right]
                 \left[1~\matrix{1&0\cr 1&0\cr}\right]
                 \left[2~\matrix{2&0\cr 2&0\cr}\right]
       ~u_i~u_0^2~\left[\matrix{v\cr v\cr} \right]$ &4~ \tabroom \\
{\rm IV}$_i$ &$ \left[0~\matrix{0&0\cr 0&0\cr}\right]
                ~g_i~\left[0~\matrix{0&0\cr 0&0\cr}\right]
                \left[0~\matrix{0&0\cr 0&0\cr}\right]
                \left[3~\matrix{3&0\cr 3&0\cr}\right]
        ~u_0~u_i~u_0~\left[\matrix{v\cr v\cr} \right]$ &4~ \tabroom \\
{\rm V}$_i$ &$ \left[0~\matrix{0&0\cr 0&0\cr}\right]
               ~g_i~\left[0~\matrix{0&0\cr 0&0\cr}\right]
               \left[1~\matrix{1&0\cr 1&0\cr}\right]
               \left[2~\matrix{2&0\cr 2&0\cr}\right]
           ~u_0~u_i~u_0~\left[\matrix{v\cr v\cr} \right]$ &4~ \tabroom \\
{\rm VI}$_i$ &$ \left[0~\matrix{0&0\cr 0&0\cr}\right]
                \left[0~\matrix{0&0\cr 0&0\cr}\right]
                ~g_i~\left[0~\matrix{0&0\cr 0&0\cr}\right]
                \left[3~\matrix{3&0\cr 3&0\cr}\right]
       ~u_0^2~u_i~\left[\matrix{v\cr v\cr} \right]$ &4~ \tabroom \\
{\rm VII}$_i$ &$ \left[0~\matrix{0&0\cr 0&0\cr}\right]
                 \left[0~\matrix{0&0\cr 0&0\cr}\right]
                ~g_i~\left[1~\matrix{1&0\cr 1&0\cr}\right]
                 \left[2~\matrix{2&0\cr 2&0\cr}\right]
        ~u_0^2~u_i~\left[\matrix{v\cr v\cr} \right]$ &4~ \tabroom \\
{\rm VIII}$_{ij}$ &$ g_i~g_j~\left[0~\matrix{0&0\cr 0&0\cr}\right]
                  \left[0~\matrix{0&0\cr 0&0\cr}\right]
                  \left[0~\matrix{0&0\cr 0&0\cr}\right]
                  \left[1~\matrix{1&0\cr 1&0\cr}\right]
         ~u_i~u_j~u_0~\left[\matrix{v\cr v\cr} \right]$ &8~ \tabroom \\
{\rm IX}$_{ij}$ &$ g_i~
                 \left[0~\matrix{0&0\cr 0&0\cr}\right]
                  ~g_j~\left[0~\matrix{0&0\cr 0&0\cr}\right]
                  \left[0~\matrix{0&0\cr 0&0\cr}\right]
                  \left[1~\matrix{1&0\cr 1&0\cr}\right]
       ~u_i~u_0~u_j~\left[\matrix{v\cr v\cr} \right]$ &8~ \tabroom \\
{\rm X}$_{ij}$  &$ \left[0~\matrix{0&0\cr 0&0\cr}\right]
                  ~g_i~g_j~\left[0~\matrix{0&0\cr 0&0\cr}\right]
                  \left[1~\matrix{1&0\cr 1&0\cr}\right]
       ~u_0~u_i~u_j~\left[\matrix{v\cr v\cr} \right]$ &8~ \tabroom \\ [2ex]
\hline
\end{tabular}
\end{center}
\end{scriptsize}
\centerline{{\bf Table 19:}~{\it Generations of the (50,0)
          SU(3)$\times $SU(2) daughter of the $3^{\otimes 5}$ model.}}
whereas the {\bf 3}s can be found in Table 20 in which we have
used the abbreviations
\bea
s_0 = \left[2~\matrix{2&0\cr -4&-2\cr}\right],&&~~~
s_1 = \left[2~\matrix{-3&-1\cr -4&-2\cr}\right], \nn \\
t_0 = \left[1~\matrix{3&2\cr -3&-2\cr}\right],&&~~~
t_1 = \left[1~\matrix{-2&-1\cr -3&-2\cr}\right].
\eea
Note that the $s_i$s are just the $v_i$s with conjugate charge on the
left side.
\pano
\begin{scriptsize}
\begin{center}
\begin{tabular}{| l r r|}
\hline
Type   &Field      &Number \tabroom \\
\hline
{\rm A}$_{ij}$  &$v_i~v_j~
           \left[0~\matrix{0&0\cr -2&-2\cr}\right]
           \left[3~\matrix{-3&0\cr -5&-2\cr}\right]
          \left[3~\matrix{-3&0\cr -5&-2\cr}\right]
         ~u_i^+~u_j^+~u_0~\left[\matrix{0\cr v\cr} \right]$ &4 \tabroom \\
{\rm B}$_{ij}$  &$v_i~
          \left[0~\matrix{0&0\cr -2&-2\cr}\right]
            ~v_j~\left[3~\matrix{-3&0\cr -5&-2\cr}\right]
          \left[3~\matrix{-3&0\cr -5&-2\cr}\right]
          ~u_i^+~u_0~u_j^+~\left[\matrix{0\cr v\cr} \right]$ &4 \tabroom \\
{\rm C}$_{ij}$  &$ \left[0~\matrix{0&0\cr -2&-2\cr}\right]
          ~v_i~v_j~\left[3~\matrix{-3&0\cr -5&-2\cr}\right]
          \left[3~\matrix{-3&0\cr -5&-2\cr}\right]
       ~u_0~u_i^+~u_j^+~\left[\matrix{0\cr v\cr} \right]$ &4 \tabroom \\
{\rm D}$_{ijk}$  &$v_i~v_j~v_k~
               \left[1~\matrix{-1&0\cr -3&-2\cr}\right]
              \left[3~\matrix{-3&0\cr -5&-2\cr}\right]
      ~u_i^+~u_j^+~u_k^+~\left[\matrix{0\cr v\cr} \right]$ &16 \tabroom \\
{\rm E}$_{ijk}$  &$v_i~v_j~v_k~
               \left[2~\matrix{-2&0\cr -4&-2\cr}\right]
              \left[2~\matrix{-2&0\cr -4&-2\cr}\right]
       ~u_i^+~u_j^+~u_k^+~\left[\matrix{0\cr v\cr} \right]$ &8 \tabroom \\
{\rm F}$_i$  &$t_i~
         \left[3~\matrix{-4&-1\cr -5&-2\cr}\right]
         \left[3~\matrix{-4&-1\cr -5&-2\cr}\right]
         \left[1~\matrix{-3&-2\cr 1&0\cr}\right]
         \left[2~\matrix{-2&0\cr 2&0\cr}\right]
         ~u_i^+~u_1^{+2}~\left[\matrix{0\cr v\cr} \right]$ &4 \tabroom \\
{\rm G}$_i$  &$\left[3~\matrix{-4&-1\cr -5&-2\cr}\right]
         ~t_i~\left[3~\matrix{-4&-1\cr -5&-2\cr}\right]
         \left[1~\matrix{-3&-2\cr 1&0\cr}\right]
         \left[2~\matrix{-2&0\cr 2&0\cr}\right]
       ~u_1^+~u_i^+~u_1^+~\left[\matrix{0\cr v\cr} \right]$ &4 \tabroom \\
{\rm H}$_i$  &$\left[3~\matrix{-4&-1\cr -5&-2\cr}\right]
         \left[3~\matrix{-4&-1\cr -5&-2\cr}\right]
         ~t_i~\left[1~\matrix{-3&-2\cr 1&0\cr}\right]
         \left[2~\matrix{-2&0\cr 2&0\cr}\right]
         ~u_1^{+2}~u_i^+~\left[\matrix{0\cr v\cr} \right]$ &4 \tabroom \\
{\rm I}$_i$  &$s_i~
         \left[3~\matrix{-4&-1\cr -5&-2\cr}\right]
         \left[3~\matrix{-4&-1\cr -5&-2\cr}\right]
         \left[2~\matrix{-2&0\cr 2&0\cr}\right]
         \left[2~\matrix{-2&0\cr 2&0\cr}\right]
        ~u_i^+~u_1^{+2}~\left[\matrix{0\cr v\cr} \right]$ &2 \tabroom \\
{\rm J}$_i$  &$ \left[3~\matrix{-4&-1\cr -5&-2\cr}\right]
         ~s_i \left[3~\matrix{-4&-1\cr -5&-2\cr}\right]
         \left[2~\matrix{-2&0\cr 2&0\cr}\right]
         \left[2~\matrix{-2&0\cr 2&0\cr}\right]
        ~u_1^+~u_i^+~u_1^+~\left[\matrix{0\cr v\cr} \right]$ &2 \tabroom \\
{\rm K}$_i$  &$ \left[3~\matrix{-4&-1\cr -5&-2\cr}\right]
         \left[3~\matrix{-4&-1\cr -5&-2\cr}\right]
         ~s_i~\left[2~\matrix{-2&0\cr 2&0\cr}\right]
         \left[2~\matrix{-2&0\cr 2&0\cr}\right]
    ~u_1^{+2}~u_i^+~\left[\matrix{0\cr v\cr} \right]$ &2 \tabroom \\ [2ex]
\hline
\end{tabular}
\end{center}
\end{scriptsize}
\centerline{{\bf Table 20:}~{\it The ${\bf 3}$s of the (50,0)
          SU(3)$\times $SU(2) daughter of the $3^{\otimes 5}$ model}}
\vskip -.1truein
\subsection{Exact Yukawa couplings}
Due to the following decomposition of the representations of E$_3$
in those of SO($4$)$\times$ U($1$)
\bea
 {\bf 6}=({\bf 3},{\bf 2})&=&{\bf 4}^v_{-1}\oplus{\bf 2}^c_2 \nn \\
 {\bf 3}=({\bf 3},{\bf 0})&=&{\bf 1}_2\oplus{\bf 2}^s_{-1}
\eea
we can calculate the Yukawa coupling in the following way
\beq
 <{\bf 3}~{\bf 6}~{\bf 6}>=<{\bf 1}_2~{\bf 4}^v_{-1}~{\bf 4}^v_{-1}>.
\eeq
Proceeding in a completely analogous way as in the two former examples
leads to the nonvanishing Yukawas of Table 21.
\pano
\begin{scriptsize}
\begin{center}
\begin{tabular}{|l| l |}
\hline
${\bf 3}$    &Generations  \tabroom \\
\hline
{\rm A}$_{ij} \cdot$
 &{\rm I}$\cdot ${\rm VIII}$_{ij} =1$,~{\rm II}$_i\cdot ${\rm IV}$_j=1$,~
                 {\rm III}$_i\cdot ${\rm V}$_j=1$  \tabroom \\
{\rm B}$_{ij} \cdot$
 &{\rm I}$\cdot ${\rm IX}$_{ij} =1$,~{\rm II}$_i\cdot ${\rm VI}$_j=1$,~
                 {\rm III}$_i\cdot ${\rm VII}$_j=1$  \tabroom \\
{\rm C}$_{ij}\cdot$
 &{\rm I}$\cdot ${\rm X}$_{ij} =1$,~{\rm IV}$_i\cdot ${\rm VI}$_j=1$,~
                 {\rm V}$_i\cdot ${\rm VII}$_j=1$  \tabroom \\
{\rm D}$_{ijk}\cdot $
 &{\rm II}$_i\cdot ${\rm X}$_{jk}=1$,~{\rm III}$_i\cdot ${\rm X}$_{jk}=1$,~
  {\rm IV}$_j\cdot ${\rm IX}$_{ik}=1$,~{\rm V}$_j\cdot {\rm IX}_{ik}=1$,~
  {\rm VI}$_{k}\cdot ${\rm VIII}$_{ij}=1$,~{\rm VII}$_{k}\cdot
       ${\rm VIII}$_{ij}=1$ \tabroom \\
{\rm E}$_{ijk}\cdot $
 &{\rm III}$_i\cdot ${\rm X}$_{jk}=\k$,~{\rm V}$_j\cdot${\rm IX}$_{ik}=\k$,
 ~{\rm VII}$_k\cdot${\rm VIII}$_{ij}=\k$,~ \tabroom\\ [1ex]
\hline
\end{tabular}
\end{center}
\end{scriptsize}
\centerline{{\bf Table 21:}~{\it Yukawa couplings of the generations of the
         SU(3)$\times $SU(2) (50,0) daughter of $3^{\otimes 5}$.}}
\vskip -.1truein
\subsection{Extracting the chiral ring}
As is familiar by now, the structure of the chiral primary ring can be
read off from the form of the generations top down. It is clear from the
structure of the simple current that the weights of the chiral fields of
the last two minimal factors in the tensor model are not changed. Thus we
are led to introduce the coordinates of Table 22.
\begin{center}
\begin{footnotesize}
\begin{tabular}{| l r|}
\hline
Coordinate      &Weight   \tab2 \\
\hline
$x_1= \left[0~\matrix{0&0\cr 0&0\cr}\right] \otimes
          \left[0~\matrix{0&0\cr 0&0\cr}\right] \otimes
        \left[0~\matrix{0&0\cr 0&0\cr}\right] \otimes
        \left[1~\matrix{1&0\cr 1&0\cr}\right] \otimes
         \left[0~\matrix{0&0\cr 0&0\cr}\right] \otimes
          \left[\matrix{0\cr 0\cr}\right] \otimes
          \left[\matrix{0\cr 0\cr}\right] \otimes
          \left[\matrix{0\cr 0\cr}\right]  $ &1 \tabroom \\ [2ex]
$x_2 = \left[0~\matrix{0&0\cr 0&0\cr}\right] \otimes
        \left[0~\matrix{0&0\cr 0&0\cr}\right] \otimes
         \left[0~\matrix{0&0\cr 0&0\cr}\right] \otimes
        \left[0~\matrix{0&0\cr 0&0\cr}\right] \otimes
         \left[1~\matrix{1&0\cr 1&0\cr}\right]\otimes
          \left[\matrix{0\cr 0\cr}\right] \otimes
          \left[\matrix{0\cr 0\cr}\right] \otimes
          \left[\matrix{0\cr 0\cr}\right] $ &1 \tabroom \\ [2ex]
\hline
$y_1=\left[2~\matrix{2&0\cr 2&0\cr}\right] \otimes
     \left[0~\matrix{0&0\cr 0&0 \cr}\right] \otimes
      \left[0~\matrix{0&0\cr 0&0\cr}\right] \otimes
        \left[0~\matrix{0&0\cr 0&0\cr}\right] \otimes
         \left[0~\matrix{0&0\cr 0&0\cr}\right]\otimes
          \left[\matrix{0\cr 0\cr}\right] \otimes
          \left[\matrix{0\cr 0\cr}\right] \otimes
          \left[\matrix{0\cr 0\cr}\right] $ &2 \tabroom \\ [2ex]
$y_2=\left[2~\matrix{-3&-1\cr 2&0\cr}\right] \otimes
     \left[0~\matrix{0&0\cr 0&0 \cr}\right] \otimes
      \left[0~\matrix{0&0\cr 0&0\cr}\right] \otimes
        \left[0~\matrix{0&0\cr 0&0\cr}\right] \otimes
         \left[0~\matrix{0&0\cr 0&0\cr}\right]\otimes
          \left[\matrix{-1\cr 0\cr}\right] \otimes
          \left[\matrix{0\cr 0\cr}\right] \otimes
          \left[\matrix{0\cr 0\cr}\right] $ &2 \tabroom \\ [2ex]
\hline
$z_1= \left[0~\matrix{0&0\cr 0&0 \cr}\right] \otimes
     \left[2~\matrix{2&0\cr 2&0\cr}\right] \otimes
      \left[0~\matrix{0&0\cr 0&0\cr}\right] \otimes
        \left[0~\matrix{0&0\cr 0&0\cr}\right] \otimes
         \left[0~\matrix{0&0\cr 0&0\cr}\right] \otimes
          \left[\matrix{0\cr 0\cr}\right] \otimes
          \left[\matrix{0\cr 0\cr}\right] \otimes
          \left[\matrix{0\cr 0\cr}\right] $ &2 \tabroom \\ [2ex]
$z_2=\left[0~\matrix{0&0\cr 0&0 \cr}\right] \otimes
      \left[2~\matrix{-3&-1\cr 2&0\cr}\right] \otimes
      \left[0~\matrix{0&0\cr 0&0\cr}\right] \otimes
        \left[0~\matrix{0&0\cr 0&0\cr}\right] \otimes
         \left[0~\matrix{0&0\cr 0&0\cr}\right]\otimes
          \left[\matrix{0\cr 0\cr}\right] \otimes
          \left[\matrix{-1\cr 0\cr}\right] \otimes
          \left[\matrix{0\cr 0\cr}\right] $ &2 \tabroom \\ [2ex]
\hline
$w_1= \left[0~\matrix{0&0\cr 0&0 \cr}\right] \otimes
     \left[0~\matrix{0&0\cr 0&0 \cr}\right] \otimes
      \left[2~\matrix{2&0\cr 2&0\cr}\right] \otimes
        \left[0~\matrix{0&0\cr 0&0\cr}\right] \otimes
         \left[0~\matrix{0&0\cr 0&0\cr}\right]\otimes
          \left[\matrix{0\cr 0\cr}\right] \otimes
          \left[\matrix{0\cr 0\cr}\right] \otimes
          \left[\matrix{0\cr 0\cr}\right] $ &2 \tabroom \\ [2ex]
$w_2= \left[0~\matrix{0&0\cr 0&0 \cr}\right] \otimes
     \left[0~\matrix{0&0\cr 0&0 \cr}\right] \otimes
      \left[2~\matrix{-3&-1\cr 2&0\cr}\right] \otimes
        \left[0~\matrix{0&0\cr 0&0\cr}\right] \otimes
         \left[0~\matrix{0&0\cr 0&0\cr}\right]\otimes
          \left[\matrix{0\cr 0\cr}\right] \otimes
          \left[\matrix{0\cr 0\cr}\right] \otimes
          \left[\matrix{-1\cr 0\cr}\right] $ &2 \tabroom \\ [2ex]
\hline
\end{tabular}
\end{footnotesize}
\end{center}
\centerline{{\bf Table 22:}~{\it Coordinates of the monomial ring for
                the (50,0) model.}}
The generations then take the form shown in Table 23.
\begin{center}
\begin{tabular}{|l c c|}
\hline
Family   &Monomial Representative   &Degeneracy   \tabroom \\
\hline
I    &~~~~  $x_i^2x_j^3$       &2 \tabroom  \\
II$_l$   &~~~~  $u_lx_i^3$         &4 \tabroom \\
III$_l$  &~~~~  $u_lx_ix_j^2$      &4 \tabroom  \\
IV$_m$   &~~~~  $v_mx_i^3$         &4 \tabroom  \\
V$_m$    &~~~~  $v_mx_ix_j^2$      &4 \tabroom  \\
VI$_n$   &~~~~  $w_nx_i^3$         &4 \tabroom  \\
VII$_n$  &~~~~  $w_nx_ix_j^2$      &4 \tabroom \\
VIII$_{lm}$ &~~~~  $u_lv_mx_i$        &8 \tabroom \\
IX$_{ln}$   &~~~~  $u_lw_nx_i$        &8 \tabroom  \\
X$_{mn}$    &~~~~  $v_mw_nx_i$        &8 \tabroom  \\ [1ex]
\hline
\end{tabular}
\end{center}
\centerline{{\bf Table 23:}~{\it Monomial representation of the generations
                  of the (50,0) model.}}
\relax From the generations we read off the ideal
\beq
\cI [x_i^4, u_au_b, v_av_b, w_aw_b],
\eeq
from which we see in turn that the monomials of degree ten
decompose into the types listed in Table 24.
\begin{center}
\begin{tabular}{|l c c |}
\hline
Family   &Monomial Representative   &Degeneracy   \tabroom \\
\hline
A$_{lm}$    &~~~~$u_lv_mx_1^3x_2^3$     &$4$\tabroom\\
B$_{ln}$    &~~~~$u_lw_nx_1^3x_2^3$     &$4$\tabroom\\
C$_{mn}$    &~~~~$v_mw_nx_1^3x_2^3$     &$4$\tabroom\\
D$_{lmn}$   &~~~~$u_lv_mw_nx_ix_j^3$    &$16$\tabroom\\
E$_{lmn}$   &~~~~$u_lv_mw_nx_i^2x_j^2$  &$8$\tabroom\\ [1ex]
\hline
\end{tabular}
\end{center}
\centerline{{\bf Table 24:}~{\it Monomial representation of the
                 {\bf 3}s of the (50,0) model.}}
Thus we find that in the present case only 36 of the total of 54 {\bf 3}s
admit a monomial representation. Comparing the product structure of the
chiral ring with the exact Yukawa couplings we again find complete
agreement using the renormalization
\beq
{\rm E} \lra \k {\rm E}.
\eeq
For this model we have the least predictive power of the three
exactly solvable theory we have discussed. As in the first two
examples the number of undetermined scalings
is given by (rk$(V) -2)$ and therefore we find that we can adjust four
different renormalizations.
\vskip -.1truein
\subsection{Remarks concerning a possible $\si$--model representation}
The story for the $\si$--model of the (50,0) model follows the pattern
developed in the previous Sections for the (80,0) and the (64,0) models.
The additional simple current has led to a further pair of coordinates
of weight two. Again the ideal is generated by monomials of weight four
and therefore we are led to consider the base space
\beq
\IP_{(1,1,2,2,2,2,2,2,5,5)}[4~4~4~4~4~4],
\eeq
where we have added two coordinates of weight five for reasons
described already in our discussion of the (64,0) model in Section 5.
Again this space does not admit transverse polynomials and will have
hypersurface singularities, which in the present example is a complex
curve, the projective line. Computing the second Chern class, we are
led to consider yet again the same vector bundle
\beq
V_{(1,1,1,1,1;5)} \lra \IP_{(1,1,2,2,2,2,2,2,5,5)}[4~4~4~4~4~4],
\eeq
which presumably is modified by the resolution of the real--dimension two
singular set in such a way as to lead to a rank six bundle.
\par
Comparing the $\si$--models of all three models discussed in this paper
the emerging pattern indicates that a hypersurface singular set
of real--dimension $d$ should increase the rank of the bundle to $(4+d)$.
\section{Conclusion}
In the present paper we have established by construction the existence
of conformal fixed points of (0,2) Calabi--Yau $\si$--models.
Along the way we have identified the exactly solvable nature of such a
conformal fixed point for the particular framework of (0,2) $\si$--models
discussed in \cite{dk94}. This generalizes to the framework of (0,2)
compactification the work of \cite{g87,lg,ew93} on
the triality of exactly solvable models, Landau--Ginzburg theories and
Calabi--Yau manifolds in the context of (2,2) compactifications. Our
result thus unifies the different description of (0,2)
vacua in a similar manner and allows us to call on many different
techniques which are available in conformal field theory,
the framework of (non)linear $\si$--models and the theory
of stable vector bundles over algebraic varieties, in order
to attack some of the outstanding questions mentioned in the
introduction, in particular possible generalizations of mirror
symmetry and the structure of the (0,2) moduli space.
\par
There are a number of avenues that present themselves for further
exploration. First, it is important to understand the map between
the chiral primary fields of the exact theory and the chiral rings
in a more systematic fashion than we have described
here. Simple currents behave like orbifolds in many respects, which
raises the hope that the analysis of \cite{ls90} can be generalized to
the framework of (0,2) theories. Even without such an understanding
it would be of interest to establish more (0,2) triality relations
even in the context of the type of exactly solvable (0,2) theories
which we have discussed. Instead of modifying Gepner models
with simple currents, one can also start with  more general classes
of N$=$2 superconformal field theories,
such as Kazama--Suzuki models and construct (0,2) models by
modifying their modular invariants with appropriate simple currents.
\vskip .2truein
\noindent
{\bf Acknowledgements:}
It is a pleasure to thank Per Berglund, Shyamoli Chaudhuri and Louise Dolan
for discussions. A.W.\ is also grateful to Werner Nahm. This work is
supported in part by U.S.\ DOE grant No.\ DE-FG05-85ER-40219 and by NSF
grant PHY--94--07194.


\begin{thebibliography}{9}
\baselineskip=16.5pt
\bibitem{bdfm88} T.\ Banks, L.\ Dixon, D.\ Friedan and E.\ Martinec,
   Nucl.\ Phys.\ {\bf B299}(1988)613
\bibitem{freeferm} H.\ Kawai, D.C.\ Lewellen and S.H.\ Tye,
   Phys.\ Rev.\ Lett.\ {\bf 57}(1986)1832,
   Nucl.\ Phys.\ {\bf B288}(1987)1;\\
 I.\ Antoniadis, C.\ Bachas and C.\ Kounnas,
   Nucl.\ Phys.\ {\bf B289}(1987)87
\bibitem{lattice} W.\ Lerche, D.\ L\"ust and A.N.\ Schellekens,
   Nucl.\ Phys.\ {\bf B287}(1987)477
\bibitem{asymorbi} K.S.\ Narain, M.H.\ Sarmadi and C.\ Vafa,
   Nucl.\ Phys.\ {\bf B288}(1987)551
\bibitem{fimqr89} A.\ Font, L.E.\ Ib\'a$\tilde{\rm n}$ez, M.\ Mondragon,
     F.\ Quevedo and G.G.\ Ross, Phys.\ Lett.\ {\bf B227}(1989)34;\\
   A.\ Font, L.E.\ Ib\'a$\tilde{\rm n}$ez, F.\ Quevedo and A.\ Sierra,
     Nucl.\ Phys.\ {\bf B337}(1990)119
\bibitem{sy90} A.N.\ Schellekens and S.\ Yankielowicz,
     Nucl.\ Phys.\ {\bf B330}(1990)103
\bibitem{bjkz95} P.\ Berglund, C.V.\ Johnson, S.\ Kachru and P.\ Zaugg,
    {\it Heterotic Coset Models and $(0,2)$ String Vacua}, hep--th/9509170
\bibitem{rs87} R.\ Schimmrigk, Phys.\ Lett.\ {\bf B193}(1987)175;\\
 D.\ Gepner, {\it String Theory on Calabi--Yau Manifolds:\ the three
 Generations Case}, preprint PUPT-88-0085, December 1987, hep--th/9301089
\bibitem{rs90} R.\ Schimmrigk, Nucl.\ Phys.\ {\bf B342}(1990)231
\bibitem{cdls88} P.\ Candelas, A.\ Dale, C.A.\ L\"utken and R.\ Schimmrigk,
                 Nucl.\ Phys.\ {\bf B298}(1988)493
\bibitem{as95} A.\ Strominger, Nucl.\ Phys.\ {\bf B451}(1995)96,
    hep--th/9504090;\\
   B.R.\ Greene, D.R.\ Morrison and A.\ Strominger, 
    Nucl.\ Phys.\ {\bf B451}(1995)109, hep--th/9504145
\bibitem{g88} D.\ Gepner, Nucl.\ Phys.\ {\bf B296}(1988)757
\bibitem{g87} D.\ Gepner, Phys.\ Lett.\ {\bf B199}(1987)380
\bibitem{lg} E.\ Martinec, Phys.\ Lett.\ {\bf B217}(1989)431;\\
  C.\ Vafa and N.P.\ Warner, Phys.\ Lett.\ {\bf B218}(1989)51;\\
  B.R.\ Greene, C.\ Vafa and N.\ Warner,
   Nucl.\ Phys.\ {\bf B324}(1989)371;\\
  C.\ Vafa, Mod.\ Phys.\ Lett.\ {\bf A4}(1989)1169
\bibitem{ew93} E.\ Witten, Nucl.\ Phys.\ {\bf B403}(1993)159, 
        hep--th/9301042
\bibitem{as90} B.\ de Wit and A.\ van Proeyen,
               Nucl.\ Phys.\ {\bf B245}(1984)89;\\
 E.\ Cremmer, C.\ Kounnas, A.\ van Proeyen, J.P.\ Derendinger,
  B.\ de Wit and L.\ Girardello, Nucl.\ Phys.\ {\bf B250}(1985)385;\\
 L.\ Dixon, V.\ Kaplunovsky and J.\ Louis,
     Nucl.\ Phys.\ {\bf B329}(1988)27;\\
      A.\ Strominger, Comm.\ Math.\ Phys.\ {\bf 133}(1990)163
\bibitem{wms} P.\ Candelas, M.\ Lynker and R.\ Schimmrigk,
      Nucl.\ Phys.\ {\bf B341}(1990)383;\\
     B.R.\ Greene and R.\ Plesser, Nucl.\ Phys.\ {\bf B338}(1990)15
\bibitem{ls90} M.\ Lynker and R.\ Schimmrigk,
     Phys.\ Lett.\ {\bf B249}(1990)237
\bibitem{sms} S.\ Kachru and C.\ Vafa, 
     Phys.\ Lett.\ {\bf B361}(1995)59, hep--th/9505105;\\
    S.\ Ferrara, J.\ Harvey, A.\ Strominger and C.\ Vafa, 
     Nucl.\ Phys.\ {\bf B450}(1995)69, hep--th/9505162
\bibitem{others} A.\ Klemm, W.\ Lerche and P.\ Mayr, 
     Phys.\ Lett.\ {\bf B357}(1995)313, hep--th/9506112;\\
    S.\ Kachru, A.\ Klemm, W.\ Lerche, P.\ Mayr and C.\ Vafa,
     {\it Nonperturbative Results on the Point Particle Limit of $N=2$
     Heterotic String Compactification}, hep--th/9508155
\bibitem{rs94} R.\ Schimmrigk, {\it Scaling Behaviour in String Theory},
         hep--th/9412077
\bibitem{ew86} E.\ Witten, Nucl.\ Phys.\ {\bf B268}(1986)79
\bibitem{g88b} D.\ Gepner, Nucl.\ Phys.\ {\bf B311}(1988)191
\bibitem{ss88} G.\ Sotkov and M.\ Stanishkov,
         Phys.\ Lett.\ {\bf B215}(1988)674
\bibitem{glr89} B.R.\ Greene, C.A.\ L\"utken and G.G.\ Ross,
         Nucl.\ Phys.\ {\bf B325}(1989)101
\bibitem{lr88} C.A.\ L\"utken and G.G.\ Ross,
         Phys.\ Lett.\ {\bf B213}(1988)152
\bibitem{ls88} M.\ Lynker and R.\ Schimmrigk,
 Phys.\ Lett.\ {\bf B215}(1988)681; Nucl.\ Phys.\ {\bf B339}(1990)121
\bibitem{fkss89} J.\ Fuchs, A.\ Klemm, C.\ Scheich and M.G.\ Schmidt,
         Phys.\ Lett.\ {\bf B232}(1989)317; Ann.\ Phys.\ {\bf 204}(1990)1
\bibitem{ks89} Y.\ Kazama and H.\ Suzuki,
 Nucl.\ Phys.\ {\bf B321}(1989)232; Phys.\ Lett.\ {\bf B216}(1989)112
\bibitem{lvw89} W.\ Lerche, C.\ Vafa and N.\ Warner,
         Nucl.\ Phys.\ {\bf B324}(1989)427
\bibitem{ls91} M.\ Lynker and R.\ Schimmrigk,
         Phys.\ Lett.\ {\bf B253}(1991)83
\bibitem{ans91} A.N.\ Schellekens, Nucl.\ Phys.\ {\bf B366}(1991)27
\bibitem{dk94} J.\ Distler and S.\ Kachru,
         Nucl.\ Phys.\ {\bf B413}(1994)213
\bibitem{kw93} S.\ Kachru and E.\ Witten,
         Nucl.\ Phys.\ {\bf B407}(1993)637, hep--th/9307038
\bibitem{dsww86} M.\ Dine, N.\ Seiberg, X.G.\ Wen and E.\ Witten,
     Nucl.\ Phys.\ {\bf B278}(1986)769; {\it ibid.\ }{\bf B289}(1987)319
\bibitem{egnq86} J.\ Ellis, C.\ Gomez, D.V.\ Nanopoulos and M.\ Quiros,
         Phys.\ Lett.\ {\bf B176}(1986)403
\bibitem{jd87} J.\ Distler, Phys.\ Lett.\ {\bf B188}(1988)431
\bibitem{dg88} J.\ Distler and B.R.\ Greene,
         Nucl.\ Phys.\ {\bf B304}(1988)1
\bibitem{dk94b} J.\ Distler and S.\ Kachru,
         Nucl.\ Phys.\ {\bf B430}(1994)13, hep--th/9406090
\bibitem{sw95} E.\ Silverstein and E.\ Witten,
         Phys.\ Lett.\ {\bf B328}(1994)307, hep--th/9403054;
         Nucl.\ Phys.\ {\bf B444}(1995)161, hep--th/9503212
\bibitem{bw95} R.\ Blumenhagen and A.\ Wi\ss kirchen, 
            Nucl.\ Phys.\ {\bf B454}(1995)561, hep--th/9506104
\bibitem{sy89} A.N.\ Schellekens and S.\ Yankielowicz,
 Nucl.\ Phys.\ {\bf B327}(1989)673; Phys.\ Lett.\ {\bf B227}(1989) 387;
 Int.\ J.\ Mod.\ Phys.\ {\bf A5}(1990)2903
\bibitem{ks94} A.\ Klemm and R.\ Schimmrigk,
        Nucl.\ Phys.\ {\bf B411}(1994)559, hep--th/9204060
\bibitem{krsk92} M.\ Kreuzer and H.\ Skarke,
        Nucl.\ Phys.\ {\bf B388}(1992)113, hep--th/9205004
\bibitem{fms86} D.\ Friedan, E.\ Martinec and S.\ Shenker,
        Nucl.\ Phys.\ {\bf B271}(1986)93
\bibitem{zf86} A.B.\ Zamolodchikov and V.A.\ Fateev,
        Sov.\ Phys.\ JETP {\bf 62}(1986)215
\bibitem{sk95} S.\ Kachru, Phys.\ Lett.\ {\bf B349}(1995)76, hep--th/9501131
\end{thebibliography}
\end{document}